\documentclass[preprint,12pt]{elsarticle}

\usepackage{graphicx}
\usepackage{dcolumn}
\usepackage{bm}
\usepackage{longtable}
\usepackage{hyperref}
\usepackage{makecell}

\usepackage[dvipsnames]{xcolor}
\usepackage{amssymb}
\usepackage{amsmath}
\usepackage{upgreek}

\journal{Ultramicroscopy}

\begin{document}

\begin{frontmatter}



\title{Neutral Helium Atom Microscopy}


\author[inst1]{Adrià Salvador Palau}

\author[inst1]{Sabrina Daniela Eder}

\affiliation[inst1]{organization={Department of Physics and Technology, University of Bergen},
            addressline={Allegaten 55}, 
            city={Bergen},
            postcode={5007}, 
            country={Norway}}
            
\affiliation[inst2]{organization={CNR-IMEM, Department of Physics, University of Genova},
            addressline={via Dodecaneso 33}, 
            city={Genova},
            postcode={16146}, 
            country={Italy}}
\author[inst2]{Gianangelo Bracco}

\author[inst1]{Bodil Holst}

\begin{abstract}
Neutral helium atom microscopy, also referred to as scanning helium microscopy and commonly abbreviated SHeM or NAM (neutral atom microscopy), is a novel imaging technique that uses a beam of neutral helium atoms as an imaging probe. The technique offers a number of advantages such as the very low energy of the incident probing atoms (less than 0.1 eV), unsurpassed surface sensitivity (no penetration into the sample bulk), a charge neutral, inert probe and a high depth of field. This opens up for a range of interesting applications such as: imaging of fragile and/or non-conducting samples without damage, inspection of 2D materials and nano-coatings, with the possibility to test properties such as grain boundaries or roughness on the Ångström scale (the wavelength of the incident helium atoms) and imaging of samples with high aspect ratios, with the potential to obtain true to scale height information of 3D surface topography with nanometer resolution: nano stereo microscopy. However, for a full exploitation of the technique, a range of experimental and theoretical issues still needs to be resolved. In this paper we review the research in the field. We do this by following the trajectory of the helium atoms step by step through the microscope: from the initial acceleration in the supersonic expansion used to generate the probing beam over the atom optical elements used to shape the beam, followed by interaction of the helium atoms with the sample (contrast properties) to the final detection and post-processing. We also review recent advances in scanning helium microscope design including a discussion of imaging with other atoms and molecules than helium.
\end{abstract}


%


\end{frontmatter}


\section{Introduction}
\label{sec:Introduction}
Neutral helium atom microscopes are surface characterisation tools that apply a beam of neutral helium atoms as imaging probe. The instruments exploit a supersonic expansion of helium gas from a high pressure reservoir (100 bar range) through a nozzle into vacuum to generate a high-intensity beam with a narrow velocity distribution, which is then collimated or focused onto a sample. The scattered intensity signal is recorded  ``point by point" and used to create an image of the sample in a manner similar to other beam probe microscopy techniques such as scanning electron microscopy or helium ion microscopy, but with the crucial difference that due to the very low energy\footnote{The energy of the helium atoms is determined by the temperature of the nozzle which can be cooled or heated. The energy range is typically between 20 meV corresponding to a de-Broglie wavelength of around 0.1 nm, and 60 meV (room temperature beam) corresponding to a wavelength of around 0.05 nm, see equation~\ref{eq:isentropic_velocity}. Helium atoms with energies in this range are referred to as thermal helium atoms, a term which is also used in this paper.} and strict surface sensitivity of the neutral helium beam, the neutral helium atoms scatter off the outermost electron density distribution of the sample. There is no penetration into the sample-material and no photons or secondary electrons are generated during the scattering process. 

Several abbreviations have been used for neutral helium microscopy over the years, i.e. NEMI for NEutral MIcroscope \cite{eder2012neutral}, HeM for Helium Microscope and NAM for Neutral Atom Microscopy \cite{witham_simple_2011}, however the community now seems to have agreed on SHeM for Scanning Helium Microscope, which was actually also one of the first abbreviations, introduced by MacLaren and Allison already in 2004~\cite{MacLaren2004}. We will use the abberaviation SHeM for the rest of this paper. 

The research behind neutral helium microscopes includes four main areas, which can be mapped to the different stages of  the imaging probe trajectory. Firstly modelling of the supersonic expansion of helium gas into vacuum, needed to establish the intensity and matter wave properties of the probing helium beam. Secondly, de-Broglie matter wave optics, which  describes the interaction of the neutral helium atoms with the optical elements, such as zone plates, pinholes and  mirrors, critical for the microscope resolution. Thirdly, helium atom surface scattering,  modelling the interaction between the neutral helium atoms and the sample, thus determining the contrast properties. Finally  the helium atom detection, targeting the difficult problem of detecting hard-to-ionise neutral helium atoms. In addition to these four area comes research specifically dedicated to the application of scanning helium microscopes. This includes problems such as optimisation of the overall configuration of the tool, advanced imaging techniques (including stereo-imaging), signal processing and image analysis.

For this review we have provided an open-source implementation of the solution to the Boltzmann equation in spherical coordinates used in previous helium microscope simulation work. It is meant as a service for those interested in pursuing their own helium microscope designs. The code can be found on GitHub: \cite{githubrepo}. We would also like to draw attention to the ray tracing simulation program of the Cambridge SHeM, provided by Lambrick and Seremet, also available on GitHuB~\cite{githubray}.

We begin this review with  a brief historical overview of the research that  made neutral helium microscopy possible. Then we follow the trajectory of a helium atom through a neutral helium microscope as described above. In this way, we review the background research, step by step. Then we move on to discuss the latest research on  microscope design and imaging techniques and we present an overview of the SHeM images published up till now. The paper finishes with an outlook on the expected future of the field.

\subsection{Some Background History}

In 1930, one year before the electron microscope was invented, Estermann and Stern scattered an effusive beam of neutral helium atoms off LiF(100) and saw diffraction peaks \cite{estermann1930beugung}.  Their groundbreaking work had been made possible thanks to previous work by Dunnoyer who established the first directed atom beam in 1911~\cite{pauly2012atom}.

The Estermann and Stern experiment did not  instigate a new research field straight away, due to limitations in  pumping technology which enabled only the production of effusive beams, which have low intensity and a broad velocity distribution (and therefore a broad de-Broglie wavelength distribution). It took another twenty years for Kantrowitz and Grey to devise a helium source with a narrower velocity distribution \cite{kantrowitz1951high}. This was achieved thanks to a supersonic expansion of helium gas into a lower-pressure chamber (see Sec. \ref{sec:the_supersonic_expansion}). Notwithstanding the clear improvement that this brought, much narrower velocity distributions and higher intensities were imperative for the success of neutral helium atoms as a scattering probe.

Such beam properties were achieved in the early 1970s thanks to the improvement of vacuum techniques and the introduction of small nozzles, which allowed for supersonic expansion into ultra high vacuum. The central part of the beam was selected using a conically shaped aperture, a so called skimmer - until that point, slits had been preferred.  By the 1980s, nozzle technology had advanced so much that the velocity distribution of the helium beams had become narrow enough that the small energy changes\footnote{meV range.} resulting from the creation or annihilation of surface phonons could be measured \cite{brusdeylins1981measurement}. This propelled  helium atom scattering as a method suitable to study surface dynamics~\cite{Benedek2018, Holst2021,bracco_surface_2013, Farias1998}.

Eventually, physicists began to speculate on how the surface sensitivity of helium could also be used to construct an imaging instrument. It soon became clear that focusing optics was a particular challenge. Neutral, ground-state helium has the smallest polarisability of all atoms and molecules. Hence manipulation via electrostatic or magnetic fields is essentially not possible unless one uses $^3He$, which is in principle possible, but up till now $^3He$ has not been applied in microscopy experiments.  Furthermore, helium atoms at thermal energies do not penetrate solid materials. In practice, the only possible way to manipulate them is via their de-Broglie matter-wave properties. This leaves only three possibilities: simple collimation, focusing via mirror reflection or focusing via diffraction from free-standing structures (zone plates).

To the best of our knowledge, the idea of a neutral helium microscope was mentioned for the first time in an official scientific context in 1990 when Doak presented results demonstrating focusing in 1D by reflecting a neutral, ground state helium beam off a mechanically bent, gold-coated piece of mica at two conferences~\cite{Doak90,Doak90-2}. Some of the results were published a couple of years later~\cite{Doak92}. A full manuscript, written at the time but never published, has been made publicly available for the first time in connection with this review~\cite{Doak-22}. In 1991 Carnal et al. presented the first experiment on 2D focusing of neutral helium beams:  The focusing of a beam of metastable helium atoms using a Fresnel zone plate~\cite{carnal1991imaging}. In 1997 Holst and Allison achieved astigmatic focusing in 2D by scattering a neutral, ground state helium beam off a Si(111)-H(1x1) surface electrostatically bent to a parabolic shape \cite{holst1997atom,holst2001}.  The silicon wafer used had a thickness of 50~$\upmu\mathrm{m}$. The area of least confusion had a spot diameter of 210~$\upmu\mathrm{m}$, which Holst et al. used in a later experiment to image the ionisation region of an electron bombardment detector~\cite{Holst1999}. In 1999 Grisenti et al. obtained the first focusing of neutral, ground state helium with a zone plate. They used a micro skimmer as a source and achieved a focused spot diameter of less than 2~$\upmu$m~\cite{doak1999towards}. 

In 1999 it was proposed that by changing the boundary conditions from round to ellipsoidal a mirror without stigmatic error could be obtained by electrostatic bending~\cite{wilson1999optical}, see also  \cite{maclaren2000single,maclaren2003,holst1999mechanical}. In 2010  Fladisher et al. achieved near stigmatic focusing of helium atoms using this method~\cite{fladischer2010ellipsoidal}. Despite work on the optimisation of the hydrogen passivation of the Si(111) surface~\cite{Maclaren2001} and the development of a transport procedure that allowed transport of mirrors to microscope systems~\cite{Maclaren2002}, it remained a problem with the Si(111)-H(1x1) mirrors that there is a considerable loss in intensity in the specular beam due to diffraction from the corrugated electron density distribution at the surface~\cite{Barredo2007,Buckland2000,Buckland1999}. In  2008 Barredo et al. showed that the reflectivity of an atom mirror could be dramatically improved by coating the silicon wafer surface with a 1-2 nm layer of lead~\cite{barredo2008quantum}. This so called-quantum stabilised mirror demonstrated a specular helium reflectivity of 67$\%$. In a later work Anemone et al. explored the use of flexible thin metal crystals as focusing mirrors \cite{anemone2017flexible}, following an early attempt from 1999 \cite{holst1999mechanical}. 
Despite the promising achievements in bent mirror focusing, the problem remains that to achieve focal spots at the nanometer range, near uniformly flat crystals without bow and warp are necessary. Work has been done  on thin wafer characterisation targeted for atom mirror applications~\cite{Litwin2005,Weeks2007,Litwin2007,Galas2007,Fladischer2008,Litwin2008}, on the improvement of the quality (flatness of a free, un-clamped wafer) of ultra-thin Si wafers for mirror applications~\cite{Sass2006} and on how wafer imperfections identified through this characterisation can be compensated for by a multiple electrode structure for bending~\cite{Ross2011}, but even so the very high technological requirements needed for nanometer range focusing, seems to have put an end to research on the creation of focusing atom mirrors through thin crystal bending, at least for the time being.

A very different approach for making atom focusing mirrors was presented by Schewe et al. in 2009~\cite{Schewe2009}. 1D focussing of a helium beam down to 1.8~$\upmu$m was demonstrated by quantum reflection from a cylindrical, concave quartz mirror. For a sufficiently small normal component of the incident wave vector of the atom, quantum reflection at the attractive branch of the helium-surface interaction potential is achieved. The great advantage of this technique is 100$\%$ reflectivity into the focus from a surface that can be microscopically rough~\cite{Zhao2010}, so that producing the mirror suddenly becomes very easy - a simple, commercial glass substrate, machined into a suitable elliposoidal shape, can be used. The problem is that near-grazing incidence is required to make the wave vector component small enough, and this puts a limit to how large a beam that can be used. 

An alternative path for  making atom focusing mirrors without thin crystal bending was proposed in 2011 by Sutter et al. who showed  that a high-reflecting mirror with a specular helium reflectivity of 23$\%$ could be obtained with a graphene-terminated Ru(0001) thin film grown on c--axis sapphire~\cite{sutter2011}. Earlier work had shown that monolayer graphene can grow on polycrystalline Ru thin films on arbitrarily shaped surfaces~\cite{Sutter2010}, this in principle, paves the way for making a focusing mirror by growing a thin layer of ruthenium on a sapphire substrate polished to the desired mirror shape and terminate it with graphene. Work pursing this is  ongoing~\cite{anemone2017}. 

\begin{figure}
    \centering
    \includegraphics[width=0.67\linewidth]{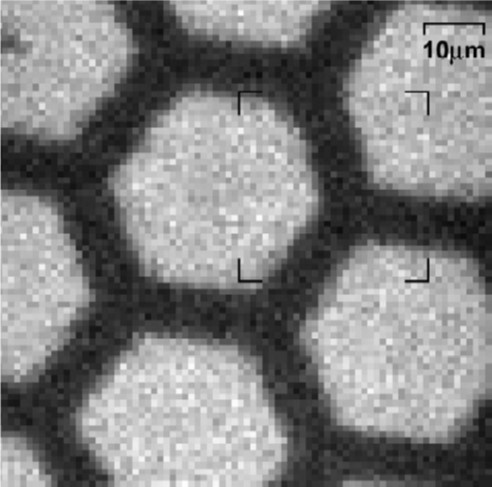}
    \caption{First SHeM image. The image is a 2D shadow image, showing a free standing grating structure. The image is obtained by scanning the focused beam across the sample, recording the signal from the transmitted helium atoms, see Fig.~\ref{zoneplate_transmission}. Image reproduced from~\cite{koch_imaging_2008}.\label{fig:firstimage}}
\end{figure}

\begin{figure}
    \centering
    \includegraphics[width=0.67\linewidth]{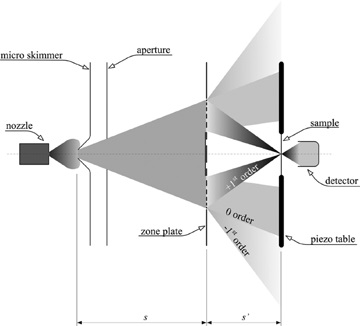}
    \caption{Diagram of the first SHeM microscope used to obtain the image shown in Fig.~\ref{fig:firstimage}. The beam is generated by a free-jet (supersonic) expansion through a micro-skimmer (see section~\ref{microskimmer}) which is imaged onto the sample plane by the zone plate's plus first diffraction order.  In addition, the zero as well as the minus first order are indicated. s is the source to zone plate distance and s' the distance from zone plate to image plane. The indicated aperture just serves to filter background from the source. The sample is scanned across the beam by a piezo table. The transmitted intensity of the beam is recorded at the back to obtain an image in transmission mode. This first SHeM microscope did not have an order-sorting aperture~\cite{eder2017zero}. Figure reproduced from~\cite{koch_imaging_2008}.\label{zoneplate_transmission}}
\end{figure}

Despite the various promising approaches, no functioning SHeM based on mirror focusing has been built so far. The first SHeM image was obtained by Koch et al. in 2008 \cite{koch_imaging_2008}, see Fig.~\ref{fig:firstimage}\footnote{For the first SHeM microscopy images Koch et al. used an instrument popularly known as MAGIE~\cite{koch_imaging_2008}. MAGIE was designed for this purpose by Graham, Holst, Toennies and technical staff at the Max Planck Institute for Fluid Mechanics in Göttingen. The instrument was built in the institute workshop. Ernst purchased  MAGIE from the Max Planck Society following the retirement of Toennies and so the experiments of Koch et al. were carried out at TU-Graz.}. Koch et al. obtained a 2D shadow image  of a free standing grating structure with a resolution of around 2~$\upmu$m using a micro-skimmer (see section~\ref{microskimmer}) and a Fresnel zone plate to focus the helium beam onto the grating. A diagram of the instrument can be seen in Fig.~\ref{zoneplate_transmission}. The best SHeM resolution obtained with a zone plate up till now is slightly less than 1~$\upmu$m, demonstrated by Eder et al.~\cite{eder2012focusing}. This is very far from the theoretical resolution limits which are discussed in section \ref{sec:resolution_limits}. Note that for the zone plate microscope configuration, the 0-order component of the beam should be blocked from entering the sample chamber in order to minimize the background signal. This can be done using a so called order-sorting aperture, also known from X-ray optics. An order sorting aperture for helium focusing with a zone plate was demonstrated in~\cite{eder2017zero}.
In 2008 a zone plate was also used by Reisinger et al. to focus a beam of Deuterium molecules, as a first demonstration of the potential of making microscopes with other atomic and molecular beams~\cite{reisinger2008focus}, see also~\cite{reisinger2012brightness}. For a description of the zone plates used for neutral helium microscopy  see~\cite{Rehbein2003,rehbein2001entwicklung,reisinger2010,reisinger2011free} .

\begin{figure}
    \centering
    \includegraphics[width=0.67\linewidth]{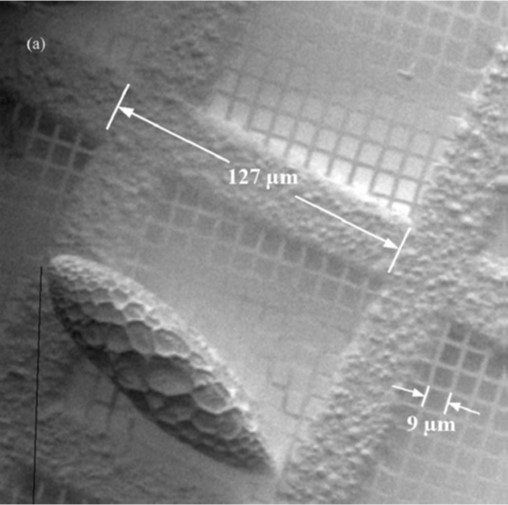}
    \caption{First SHeM image obtained in reflection mode. The image shows an uncoated pollen grain on a Quantifoil grid. The image is created by detecting the atoms scattered over a particular range of angles,  image reproduced from~\cite{witham_simple_2011}.\label{fig:secondimage}}
\end{figure}

\begin{figure}
    \centering
    \includegraphics[width=0.67\linewidth]{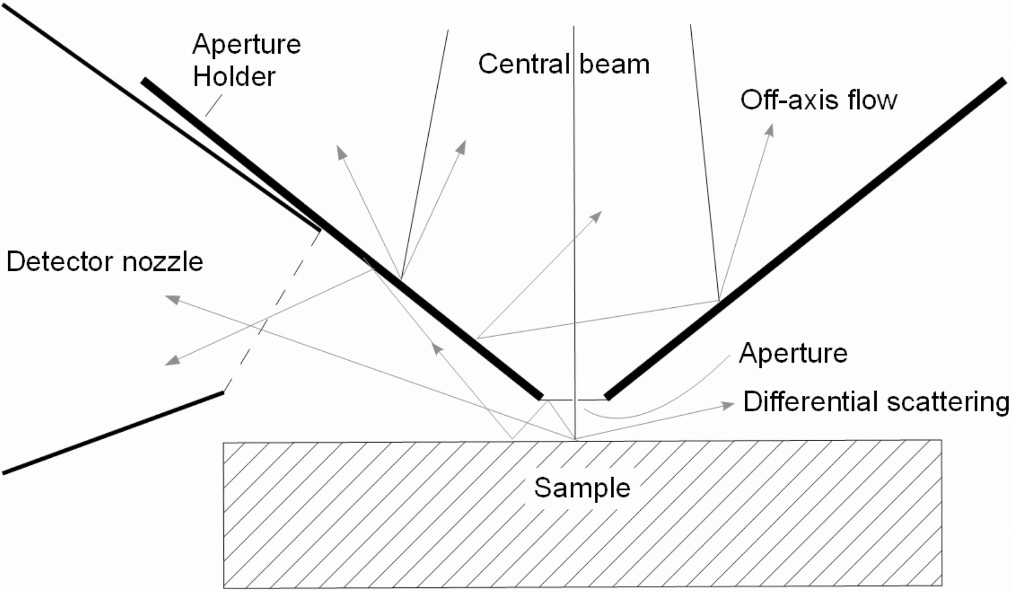}
    \caption{Diagram of the first pinhole SHeM. Figure reproduced from~\cite{witham_simple_2011} \label{fig:withams_pinhole_setup}}
\end{figure}

After the work of Koch et al., other research groups focused on a simpler configuration: the pinhole microscope. This configuration uses a small circular aperture (a pinhole) to collimate the beam instead of focusing optics. The resolution is determined by the size of the pinhole, see Fig.~\ref{fig:withams_pinhole_setup}. Using a configuration with a pinhole placed directly in front of the nozzle, Witham and Sánchez managed to obtain the first SHeM images in reflection mode in 2011 \cite{witham_simple_2011}, see Fig.~\ref{fig:secondimage}. The initial resolution was $1.5~\upmu$m and later $0.315$~$\upmu$m \cite{witham2012increased}, which remains the highest resolution obtained so far with a neutral helium atom microscope. In 2014 Witham and Sánchez also demonstrated reflection imaging with a Krypton beam \cite{witham_exploring_2014}, see also~\cite{Witham-thesis}.

Around the same time as Witham and  Sánchez other researchers had started working on pinhole microscopes with a different design, using a skimmer in combination with a collimating pinhole aperture~\cite{barr2012desktop}. Their approach was inspired  by a set of unpublished reports on neutral helium atom microscopy, written by Lower around 1992~\cite{ellis2022}. In connection with this review these reports have now been made publicly available for the first time~\cite{Lower-22-1,Lower-22-2,Lower-22-3,Lower-22-4}. The idea of a pinhole microscope is also discussed in~\cite{hustler2008aspects}. In the reports Lower discusses helium microscopy based on focusing with mirrors and zoneplates and he also introduces the idea of a microscope based purely on collimation which he calls a pinhole microscope\footnote{Before the publication of this review, Lower's reports were only known to a limited number of people with contact to the Toennies group, where Lower did the work.~\cite{hustler2008aspects} is a PhD thesis also not easily accessible. Thus,  Witham and Sánchez came up with the idea and the term pinhole microscope independently.}.

\begin{figure}
    \centering
    \includegraphics[width=0.67\linewidth]{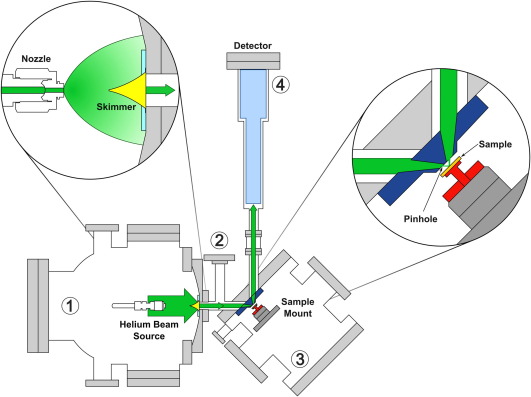}
    \caption{Schematic diagram of the second pinhole SHeM using a skimmer in combination with a pinhole. The helium beam is generated in a free-jet (supersonic) expansion in the source chamber (1), passing through a differential pumping stage (2) to the pinhole optics. The collimated beam hits the sample in the sample chamber (3). The scattered helium entering the detector chamber (4) where it stagnates to form a stable pressure, which is measured. The image is produced by scanning the sample under the beam. Figure reproduced from~\cite{barr2014design} \label{fig:dastoor_pinhole_setup}}
\end{figure}

The first images from a pinhole microscope of the type originally proposed by Lower were published in 2014 \cite{barr2014design}. For a diagram of the setup, see Fig.~\ref{fig:dastoor_pinhole_setup}. The advantage of using a skimmer is that the perturbation of the helium atoms trajectories through backscattering is reduced, see for example~\cite{palau2018center} and  section~\ref{sec:skimmer_interference}. This means that a pinhole microscope with a skimmer should in principle have an increased intensity in the beam spot on the sample and provide a narrow, well defined velocity distribution. The latter is of particular interest for contrast properties, see section~\ref{sec:contrast_properties}. A counter argument in favour of the Witham-Sánchez design is that here the pinhole can be brought closer to the nozzle, which in principle should also increase the intensity for a given working distance (distance between pinhole and sample). No detailed performance comparisons between the two designs have been presented in the literature up till now. At present it seems that the community mainly pursues the second design, where a skimmer is used. This including the most resent instrument by Bhardwaj et al.~\cite{bhardwaj2021imaging}. To this day, the pinhole microscopes remains the most widespread neutral helium microscopy design, despite the fact that higher resolution can in principle be achieved with the zone plate configuration for a given working distance (distance between optical element and sample). This is discussed in section \ref{sec:optimal_config}.

\section{The Helium Source}\label{sec:helium_source}
In a SHeM atoms start their journey in the source. A typical SHeM source follows the design established for helium atom scattering (HAS) \cite{eder2018velocity,barr2012desktop,barr2014design}:  helium is accelerated in a supersonic expansion from a high pressure reservoir, through a nozzle, into a vacuum chamber, known as the expansion chamber \cite{beijerinck1981absolute}. There, the central part of the beam is selected by a conically shaped aperture, called the skimmer\footnote{The Whitham-Sánchez pinhole microscope design skip the skimmer altogether and use a single collimating aperture downstream. \cite{witham_simple_2011}}. 

\begin{figure}
    \centering
    \includegraphics[width=0.67\linewidth]{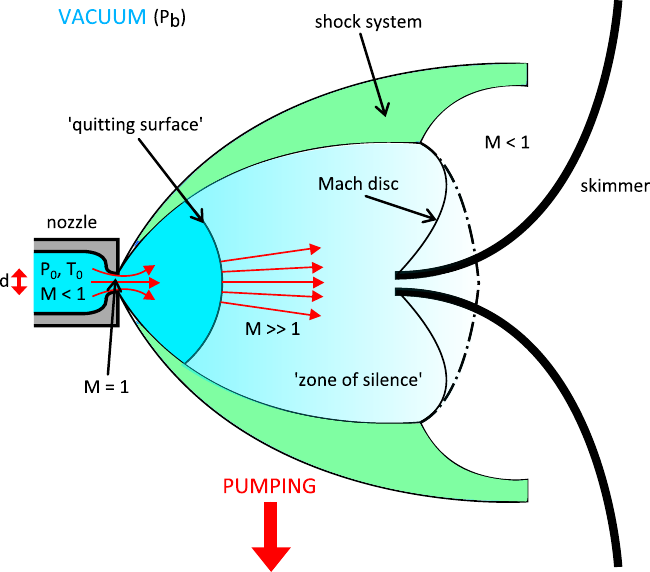}
    \caption{Schematic diagram of a free-jet (supersonic) expansion. The source expands from a helium reservoir with pressure $P_0$ and temperature $T_0$ through a nozzle of diameter $d$ into a vacuum. The Mach number $M$ (ratio of flow velocity to the local speed of sound) rapidly increases during the initial expansion. As the expansion continues there is a transition from continuum flow to molecular flow, which is often modelled with the so-called "quitting surface" (see Fig.~\ref{fig:VirtualSource}). The central part of the beam is sampled by a skimmer. In modern, high-efficient pumping systems the position of the Mach disc is often behind the skimmer and for low background pressure the shock structure and Mach disc are less pronounced and disappear. Figure reproduced from~\cite{barr2012desktop}\label{fig:Freejet}}
\end{figure}

When designing a Helium Source, one can essentially choose five parameters: temperature, pressure, nozzle radius\footnote{The issue of nozzle design is left as outside of the scope of this paper. For a discussion of this topic see for example \cite{braun1997micrometer,beijerinck1981absolute}.}, skimmer diameter and distance between skimmer and nozzle\footnote{There are also important considerations that needs to be done regarding vacuum chamber geometry, required pumping speed, shape of the skimmer etc., but that is 
beyond the scope of this paper}.  In general, small nozzles and high pressures produce brighter\footnote{Count rate per steradian and unit area of the source.} sources and therefore are more efficient in reducing undesired effects such as back-scattering interference (for a given flow, the beam is more directed) \cite{deponte2006brightness,hedgeland2005anomalous}. Similarly, cold sources are more intense than warm sources and produce higher parallel speed ratios\footnote{The speed ratio of a supersonic molecular beam is defined as $\frac{\bar v}{\Delta v}$ where $\bar v$ is the most probable velocity and $\Delta v$ is the Full Width at Half Maximum of the velocity distribution.}  \cite{eder2018velocity,palau2018center}, which allows them to reach higher centre-line intensities. The absolute differences between a cold (liquid nitrogen cooled) and a warm source at the same pressure can easily be on the order of $1\cdot 10^{13}\mathrm{counts}/s\cdot m^2$ at a distance of 2.44 m \cite{palau2018center}.


Once the nozzle size and temperature have been chosen, obtaining the beam properties amounts to (1) solving the supersonic expansion of the Helium gas into vacuum, and (2) calculating the beam intensity after the initial expansion. This chapter is structured with these two steps in mind: first, we discuss the work done on describing the supersonic expansion, and then we discuss the different models that give the beam intensity  downstream.

\subsection{The supersonic or free-jet expansion}\label{sec:the_supersonic_expansion}
As mentioned above the first component of a helium source is the nozzle, where atoms are accelerated to supersonic speeds in a physical phenomenon known as a supersonic expansion.

The theory describing supersonic expansions, also sometimes referred to as free jet expansions, was developed in the 1970s and 1980s, and is based on splitting the expansion into two regimes: the first regime, within the nozzle, follows a Navier-Stokes flow, and is solved through the isentropic nozzle model \cite{pauly2012atom,millerbook}. The second regime, from the nozzle exit onward, is modelled through the Boltzmann equation. The flow is obtained either by solving the corresponding integrals under simplifying assumptions \cite{sikora1974analysis,bossel1974skimming} or using Direct Simulation Monte Carlo (DSMC). \cite{bird1994molecular,bird1970direct,bird1976transition}. 

\subsubsection{The isentropic nozzle model}
Within the nozzle, the helium gas density is high (typically up to 200~bar) and the flow is modelled with Navier Stokes equations. The isentropic nozzle model gives the total flux per unit time (from now on, centre line intensity) stemming from a de Laval nozzle\footnote{de Laval refers to the nozzle profile. In practice most groups simply use a small hole (typically a commerical electron microscope aperture).} (assuming that the nozzle is cut-off in the sonic plane). This derivation considers an ideal gas in which the flow can be assumed to be a reversible and adiabatic process.  Therefore the gas can be considered isentropic - which means that the following analytical equation of the intensity can be obtained \cite{pauly2012atom}.
\begin{equation}\label{eq:intensity_nozzle}
I_0=\frac{P_0}{k_\mathrm{B}T_0}\sqrt{\frac{2k_\mathrm{B}T_0}{m}}\left({\pi} r_\mathrm{nz}^2\right)\sqrt{\frac{\gamma}{\gamma+1}}\left(\frac{2}{\gamma+1} \right)^{1/(\gamma-1)},
\end{equation}
where $T_0$, $P_0$ are the temperature and the pressure in the source. $r_\mathrm{nz}$ is the radius of the nozzle and $m$ is the mass of a helium atom. $\gamma$ is the heat capacity ratio ($\gamma = 5/3$ for helium), and $k_B$ is the Boltzmann constant.
One can also obtain the terminal velocity, $\bar{v}$, which can be used to provide the most probable de-Broglie wavelength of the atoms in the beam: \cite{pauly2012atom}:


\begin{equation}\label{eq:isentropic_velocity}
    \bar{v}=\sqrt{\frac{5k_B T_0}{m}}.
\end{equation}

This model is used to calculate the total flow from the nozzle - as it is well known that helium is the closest we get to an ideal gas \cite{verheijen1984quantitative}. Some groups also choose to add a correction given by the thickness of the boundary layer in a real gas. Beijerink and Verster provide a correction factor for a monoatomic gas \cite{beijerinck1981absolute}. 
 To our knowledge, all helium microscopy papers modelling the intensity of the helium beam use an initial intensity derived from the isentropic nozzle model (see Sec. \ref{sec:simplifiedimodels} for a breakdown).

\subsubsection{Post-nozzle flow}
Once the helium atoms have left the nozzle, the pressure drops and the flow is governed by the Boltzmann equation - as the Navier Stokes equations cease to apply. There are two main methods to solve the flow: either by numerically solving the Boltzmann equation under stringent assumptions  \cite{ashkenas1966proceedings,hamel1966kinetic,pauly2012atom} or by simulating the particle flow using DSMC. The latter method is more computationally intensive, but also more accurate than the former as it relies on fewer assumptions: For the first method, one typically assumes that the nozzle is a point source \cite{sikora1974analysis}. This assumption is grounded on work from Sherman and Ashkenas, which showed that a few nozzle diameters downstream, free jet streamlines become straight and can be extrapolated to a single point of origin close to the nozzle \cite{ashkenas1966proceedings,hamel1966kinetic}. The flow then can be solved using the collision integral for particles following Bose-Einstein statistics. The isentropic nozzle model at a short distance from the nozzle is used to obtain the initial conditions to start the integration\footnote{This is a rather arbitrary distance that must be large enough to guarantee spherical symmetry and small enough to satisfy equilibrium conditions, typically a few nozzle diameters.}. To solve this equation, a velocity distribution, and an interaction potential have to be assumed. The equations needed to solve the expansion are included in \cite{palau2018center,pauly2012atom}. As mentioned above for this review, we provide an open-source implementation of the solution to the Boltzmann equation in spherical coordinates \cite{githubrepo}.



The velocity distribution of the atoms is taken to be an ellipsoidal Maxwellian:
\begin{equation}\label{eq:ellipsoidal_vel_distribution}
f_\mathrm{ell}\left(\vec{v} \right)=n\left( \frac{m}{2\pi k_\mathrm{B}T_{\vert \vert}}\right)^{\frac{1}{2}}\left( \frac{m}{2\pi k_\mathrm{B}T_{\bot}}\right)\cdot
\mathrm{exp}\left( -\frac{m}{2k_\mathrm{B}T_{\vert \vert}}(v_{\vert\vert}-\bar{v})^2-\frac{m}{2k_\mathrm{B}T_{\bot}}v_{\bot}^2 \right).
\end{equation}

The choice of an ellipsoidal Maxwellian velocity distribution forms the basis to solve the spherically symmetrical Boltzmann equation \cite{grundy1969axially}. In these models, the expansion's macroscopic properties are expressed in a spherical coordinate system. The temperature is split in two terms, modelling the velocity distributions of the radial and angular component of the velocity in spherical coordinates $v_\parallel$ and $v_\bot$: $T_{\vert\vert}$ and $T_\bot$. These are proportional to the variance of the velocity in that coordinate system, for example $T_{\vert\vert} = \frac{m}{k_\mathrm{B}}\langle{(v_{\vert\vert}-v_{\vert\vert}^0)^2}\rangle$, where $v_{\vert\vert}^0$ is the parallel component of the mean velocity vector, $\bar{v}$ is the most probable velocity of the beam along the radial direction and $n$ is the number density of atoms.

On top of the assumption regarding the velocity distribution of the atoms, an interaction potential must be assumed. There are several options for this potential: the Lennard-Jones potential \cite{jones1924determination},  the Tang, Toennies and Yu (TTY), and the Hurly Moldover (HM) potentials \cite{tang1995accurate,hurly2000ab} being amongst the best known. Results of previous calculations show that the Lennard-Jones potential is accurate for source temperatures as low as 80 K \cite{pedemonte1999theoretical,pedemonte2003study}. Therefore, this is often the preferred choice by practitioners in the field as the Helium source is rarely cooled below this temperature \cite{pedemonte1999theoretical,pedemonte2003study,reisinger2007direct}. A detailed description of the Lennard-Jones potential and its implementation in the Boltzmann equation can be found in \cite{reisinger2007direct}.

The numerical solution of the Boltzmann equation in its spherical approximation provides the evolution of the average gas velocity, and the temperatures  $T_{\vert\vert}$ and $T_\bot$ with respect to the distance from the nozzle. This solution can then be used to determine the intensity of the beam at the sample plane by means of the so called quitting surface model - see Sec. \ref{sec:quitting_surface_model}. This solution can also be used to obtain the velocity distribution and speed ratio of the beam. These have been shown to be in good agreement with experimental data \cite{eder2018velocity}.

As mentioned above an alternative way of solving the Boltzmann equation, requiring less assumptions, is to directly simulate particle-to-particle interactions using DSMC \cite{bird1970direct,bird1994molecular}. This method addresses the numerical infeasibility of simulating the flow particle by particle by grouping those particles onto pseudo-molecules that are taken to represent a larger group of real molecules. DSMC requires assumptions on the interaction of the pseudo-molecules with different materials and with each other. These are normally phenomenological models such as the hard sphere model \cite{bird1994molecular}, the variable hard sphere model \cite{nanbu1990variable} and others \cite{bird1998recent}. DSMC is truer to nature than solving the Boltzmann equation under stringent assumptions but is also much more computationally expensive. Several papers have used this method to understand the behaviour of the helium expansion \cite{luria2011generation,bird1976transition,markelov2001comparative}.

\subsection{Intensity after the initial expansion}\label{sec:intensity_after_exp}
As the helium atom travels further away from the nozzle, it interacts less and less with neighbouring atoms. This means that modelling the supersonic expansion all the way to the sample plane is numerically inefficient.

Therefore, theorists often choose to use the fact that the Knudsen number of the flow increases with distance to the source, and that quasi-molecular flow is often reached before the first optical element, usually the skimmer, to build simplified models of the intensity. Quasi-molecular flow allows for the recovery of analytical expressions of the centre-line intensity, as particles can be assumed to travel in a straight line without further interactions.

Over the years, several intensity models have been proposed for helium sources. A combination of arbitrary variable labelling, numerical simplifications and empirical formulae has left researchers with no unified intensity models. The landscape is confusing, and in this paper we make an attempt to unify and simplify the different intensity models and explain how they compare with each other. 

We propose the following geometrical conventions: consider always an expansion stemming from a nozzle, followed by a skimmer. The skimmer is placed at a distance $x_S$ from the nozzle. Take $a$ as the distance between the skimmer and the axial point in which the intensity is measured. The distance between the nozzle and the measuring point is then ($x_S+a$). All the rest of physical variables correspond to those presented in Section \ref{sec:quitting_surface_model}. 

We propose that the intensity should always be given as particles per second per unit area\footnote{We chose unit area over steradians to signify the departure from spherical symmetry typical of supersonic beams.}.Usually the intensity can be assumed to be slowly-varying enough that to obtain a total intensity hitting a detector it is enough to multiply the centre-line intensity expressed as particles per second per unit area, by the detector's area. The medley of analytical formulas found in literature can be confusing, but one sees that they fall into three groups:

i)  those that treat the nozzle as a source of a spherically symmetric flux, and account any excess intensity by using an empirical factor \cite{beijerinck1981absolute}, ii) those that on top of this consider the thermal properties of the supersonic expansion through a dependency on the beam's speed ratio, and iii) those that explicitly integrate the quitting surface\footnote{Or an equivalent concept - known as the virtual source \cite{deponte2006brightness}} under some assumptions. All three families of intensity models are limited as they rely on overly simplistic assumptions, but they are useful in that they provide an analytical expression for the intensity. We start in the next section by considering the third family.

\subsubsection{The quitting surface intensity model}\label{sec:quitting_surface_model}

\begin{figure}
    \centering
    \includegraphics[width=0.67\linewidth]{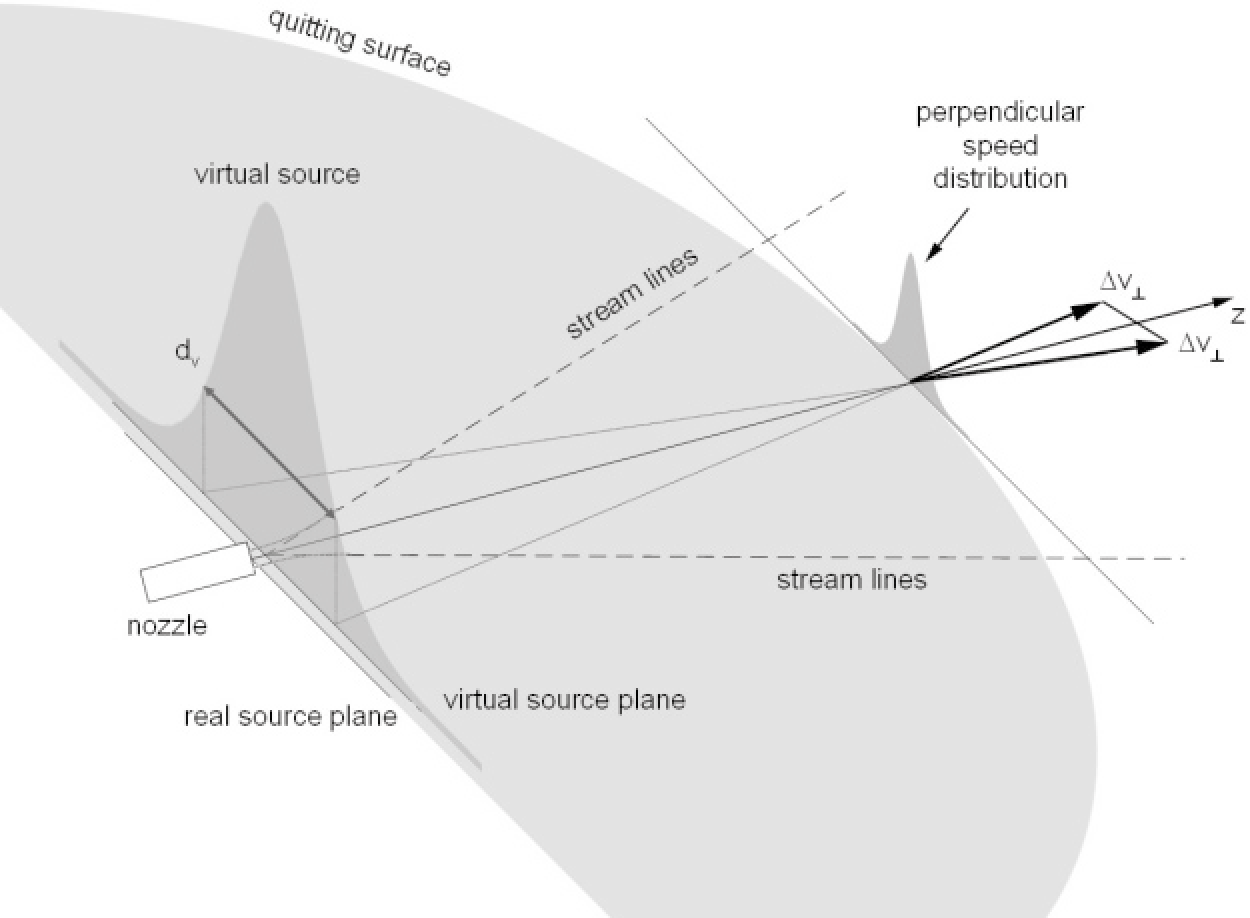}
    \caption{Schematic diagram of a supersonic expansion and definition of the virtual source. Figure reproduced from~\cite{reisinger2007direct}\label{fig:VirtualSource}}
\end{figure}

One of the most popular intensity models relies on the quitting surface model \cite{sikora1974analysis,Salvador_theoretical_2017,palau2018center} with the associated definition "virtual source". A quitting surface\footnote{Also referred to as ``last collision surface" \cite{beijerinck1981absolute,habets1977supersonic}.} is a useful theoretical construct which assumes that at a given point in the beam's supersonic expansion, particles start travelling in straight lines. This point is defined through asymptotic conditions on the properties of the expansion; either as the point in which the Mach number\footnote{Ratio of flow velocity to the local speed of sound, see \cite{hamel1966kinetic} for a discussion in the context of atom beams.} of the expansion approaches its predicted terminal Mach number \cite{witham_simple_2011,amirav1980cooling} or as the point in which the parallel and perpendicular temperature of the Maxwellian distribution used to model the expansion decouple \cite{toennies1977theoretical}.

The main utility of this model is that the intensity and velocity distribution of the beam can be obtained by integrating this spherical particle-emitting surface. This distribution can then be backtraced from the quitting surface to a so called virtual source plane, which describes the intensity and velocity distribution that a source would need to have to give rise to the observed distribution at the quitting surface. The virtual source plane is taken as the plane where the spatial distribution has the minimum extension. This means that the virtual source can be seen as the object that, with a view reduced by the skimmer, is imaged onto the sample plane by the zone plate in the zone plate microscope. Zone plates have actually been used in combination with large skimmers to obtain direct images of the supersonic expansion~\cite{reisinger2007direct,reisinger2012brightness,eder2013}.  The difficulty associated with this method is the relative arbitrarity of the definition of the quitting surface, which depending on the condition chosen, can be positioned before or after the skimmer aperture. 

As mentioned above, the quitting surface can be integrated to obtain an analytical model for the beam intensity, this is known as the Sikora approximation \cite{sikora1974analysis}. This expression was initially calculated for a quitting surface placed exactly at the skimmer aperture and was later generalised by Bossel to incorporate a quitting surface placed before the skimmer \cite{bossel1974skimming}. This formula was used, quite successfully to model experimental measurements of centre line intensities \cite{palau2018center} and is used in the first two papers by Salvador et al. on optimisation of pinhole and single zone plate helium microscope configurations~\cite{palau_theoretical_2016,Salvador_theoretical_2017}:

\begin{equation}\label{eq:I_sikora}
I_\mathrm{S}=I_{TG}\Bigg\{ 1-\exp\left[-S_i^2\left(\frac{r_\mathrm{S}(R_\mathrm{F}+a)}{R_\mathrm{F}(R_\mathrm{F}-x_\mathrm{S}+a)}\right)^2\right]\Bigg\},
\end{equation}
$I_S$ is the intensity arriving at a detector of a given radius, $r_D$, downstream, the subscript ``S" is included to indicate that the Sikora approximation is being used. $x_S$ is the distance between the nozzle and the skimmer and $a$ is the distance between the skimmer and the point where the intensity is measured. $r_s$ is the radius of the skimmer and $R_F$ is the radius of the quitting surface. The first term of the Sikora approximation $I_{TG}$ is the intensity corresponding to a naive spherically-symmetric model of the supersonic expansion, where the atoms would travel in straight lines from the nozzle with equal probability at any angle and no thermal effects\footnote{The density at the skimmer can also be used, if the point source assumption is dropped \cite{palau2018center}.}. The radius of the detector, $r_D$, is included in this term. The total flow stemming from this ideal point source corresponds to the intensity resulting from the isentropic source model $I_0$. We name this factor the thermal-geometrical component, $I_{TG}$. The thermal-geometrical intensity measured at a detector of radius $r_D$ at a distance $a$ from the skimmer is then:

\begin{equation}\label{thermal-geometrical}
    I_{TG}=\pi I_0 \frac{r_D^2}{(x_S+a)^2}.
\end{equation}

This component suffices to understand a basic design principle of neutral helium microscopy: reducing the axial length of the microscope is often beneficial, as intensity will decrease with distance.


The second term of the Sikora equation, the exponential term, models what we may term the thermal properties of the beam. The $S_i$ indicates that depending on the microscope design the perpendicular ($\bot$) or the parallel ($\lvert\lvert$) speed ratio dominates:
\begin{equation}
S_i = \sqrt{\frac{m\bar{v}^2}{2kT_{i}}},\qquad i=\lvert\lvert, \bot.
\end{equation}
Here, $T_i$ corresponds to the perpendicular or parallel temperature (as defined in eq. (\ref{eq:ellipsoidal_vel_distribution})). 

The fact that both the perpendicular and parallel speed ratios can be used in the same model of the beam intensity can be confusing. The reason behind this is that depending on where the quitting surface is assumed, one obtains an integral mostly dominated by the perpendicular speed ratio or by the parallel speed ratio - and therefore, different approximations apply. This was first demonstrated in Sikora's PhD thesis \cite{sikora1974analysis}. The rule of thumb is the following: if there are reasons to assume that the quitting surface - understood as the position of last He-He collisions - is placed very close to the skimmer, the perpendicular speed ratio should be used and $R_F=x_S$. If the quitting surface is placed far before the skimmer then the parallel speed ratio should be used. The full derivation for this can be found in the Appendix B of Sikora's PhD thesis.

Recent experimental findings show a more complex picture: for small skimmers close to the quitting surface, using the parallel speed ratio in the Sikora model reproduces experimental measurements better (as perpendicular spread is not a big contributor given that very little of the quitting surface is seen at the detector). However, for larger skimmers seeing a thermalised portion of the expansion (for example, when the expansion is not assumed to end until significantly after the skimmer) the perpendicular speed ratio dominates as predicted by Sikora \cite{palau2018center}.

As mentioned above, the thermal-geometrical term $I_{TG}$ describes a spherically symmetrical expansion. In reality as supersonically expanded atom beam decreases in intensity at a slower rate than a spherically symmetrical expansion. Thus the full Sikora-Bossel equation (eq.~\ref{eq:I_sikora}) gives a truer description of the phenomena at play. 
Let us look at this equation in the small skimmer limit: 

\begin{equation}
    I_S = I_{TG}\Bigg\{S_i^2\left(\frac{r_\mathrm{S}(R_\mathrm{F}+a)}{R_\mathrm{F}(R_\mathrm{F}-x_\mathrm{S}+a)}\right)^2\Bigg\} 
\end{equation}
For a quitting surface at the skimmer $R_F=x_S$ we get:

\begin{equation}
        I_S = I_{TG} \Bigg\{S_i^2\left(\frac{r_\mathrm{S}(x_S+a)}{a x_S}\right)^2\Bigg\}=I_{TG} \Bigg\{S_i^2\left(\frac{r_\mathrm{S}}{a}+\frac{r_\mathrm{S}}{x_S}\right)^2\Bigg\}
     \label{small-skimmer}
\end{equation}
This equation adds three important, physical corrections to the simple, spherical expansion expression ($I_{TG}$): i) the beam will be more intense the wider the skimmer is - which accounts for thermal components of the quitting surface. (ii) Higher speed ratios means more intense beams - which is a measure of the quality of the supersonic expansion and of its departure from a spherically symmetrical expansion. (iii) the closer you are to the beam source, the more intense the beam will be. Note that (i) is still an approximation and only holds for small skimmers - if the skimmer size is on the order of the size of the quitting surface, increasing it further does not result in important intensity changes. The fact that the beam intensity decreases slower than in the spherical case, has design implications. It allows for larger and hence technologically more feasible microscopes. It is important to be aware of this when doing SHeM designs (see section \ref{sec:optimal_config}).
\subsubsection{Other intensity models} 
\label{sec:simplifiedimodels}
Rather than integrating over the quitting surface/virtual source as described in the previous section   one can obtain analytical expressions for the centre line intensity by using various additional approximations. 
In this section we review other intensity models that have been used for SHeM and show how they compare to the isentropic, spherically symmetric expansion model, expressed by the thermal-geometrical term, $I_{TG}$ (see eq.~\ref{thermal-geometrical}).  The models that have been used for SHeM so far, have been inspired by centre line intensity derivations of Pauly \cite{pauly2012atom}, Miller \cite{millerbook} and dePonte \cite{deponte2006brightness} as well as Sikora (described above)

We start with models from the first family as described above, to which both Pauly and Miller belongs. These models approximate the intensity by considering a simple flow pattern: an isotropic\footnote{Note that isotropic is not the same as isentropic. Isotropic refers to the intensity being independent of the direction (spherically symmetric). Isentropic refers to the thermodynamic properties of the flow within the nozzle.} spherically symmetric expansion - such as what is assumed to obtain $I_{TG}$. The  expression for the centre line intensity presented below is used to describe SHeM among others by \cite{eder2012neutral,kaltenbacher2016optimization}. Note that the expression differs from Pauly's calculation by a factor of approximately 2. We think that this is either because of a typo - the original Miller and Pauly's equations are identical. Or because they implicitly introduce the peaking factor that Miller introduces as being approximately 2 for helium \cite{millerbook}, see also  \cite{beijerinck1981absolute}.
\begin{equation}
    I = 0.155\frac{P_0}{k_B T_0}(\frac{2r_{nz}}{x_S+a})^2\pi r_{D}^2 \sqrt{\frac{5k_B T_0}{m}}=0.62\sqrt{5/2}\frac{I_0}{f(\gamma)} \frac{r_D^2}{(x_S+a)^2}=0.6077 I_{TG}
\end{equation}
Where $f(\gamma)=\sqrt{\frac{\gamma}{\gamma+1}}\left(\frac{2}{\gamma+1} \right)^{1/(\gamma-1)}\approx  0.5135$. $\gamma=c_p/c_v$ is the heat capacity ratio, which is 5/3 for monoatomic gases and $r_{nz}$ is the radius of the nozzle.

Secondly, we look at the expression used by Witham and Sanchez \cite{witham_simple_2011} to estimate the intensity in their pinhole microscope (see Fig. \ref{fig:withams_pinhole_setup}).  Witham and Sanchez explicitly refer to Miller's derivation \cite{millerbook} for an isentropic intensity and they introduce a peaking factor $\kappa$ in their calculations.

\begin{multline}\label{eq:peaking_factor}
    I = \kappa \frac{P_0 \pi r_{nz}^2(\frac{\gamma-1}{2}+1)^{\gamma/(\gamma-1)}\sqrt{\frac{\gamma k_B T_0}{m}}}{kT_0} \pi \frac{r_D^2}{(x_S+a)^2}\\ \approx 0.4871* 2\kappa \sqrt{\gamma/2}\frac{I_0}{f(\gamma)} \frac{r_D^2}{(x_S+a)^2}\approx 1.1026 I_{TG}.
\end{multline}
 
Note that both these equations do not include the skimmer radius. In the case of Witham and Sanchez this makes sense since they are modelling a microscope design that does not include a skimmer, however, experiments have shown that in systems with a skimmer, the size of the skimmer has to be considered~\cite{palau2018center}.

Finally in~\cite{BERGIN2019112833} Bergin et al. model the source in a helium microscope using  DePonte et al's centre-line beam intensity (with a correction) \cite{deponte2006brightness,BERGIN2019112833}. This is a model of the second type, in which an empirical formula for the dependency between the virtual source radius and the speed ratio of the beam is used (and therefore an inverse dependency on the speed ratio is introduced). Bergin et al. use the following formula (in flux per unit area):

\begin{equation}
    I_{berg}=\pi\beta^2 B = \pi \left(\frac{r_S}{x_S+a}\right)^2 B.
\end{equation}
Where $B$ is the brightness of the source (in number of atoms per steradian per unit area of the source). Bergin et al. provide an expression for $B$:
\begin{equation}
    B=0.18\frac{P_0}{S_{\lvert\lvert}\sqrt{m k_B T_0}}.
\end{equation}

Combining both equations Bergin et al. arrive to a similar quadratic dependency with the skimmer radius as Sikora does in the limit of small skimmers (see eq.~\ref{small-skimmer}). Once rewritten in terms of $I_{TG}$ and multiplied by $\pi r_D^2$ the intensity arriving at a detector downstream is recovered:
\begin{equation}\label{eq:bergin_int}
    I = \frac{0.18 \pi ^2 P_0r_{S}^2}{\sqrt{m k_B T_0}}\frac{r_{D}^2}{S_{\lvert \lvert}(x_S+a)^2} =\frac{1}{S_{\lvert \lvert}}\frac{0.18\pi}{\sqrt{2}f(\gamma)} \frac{I_0}{r_{nz}^2}\frac{r_S^2r_{D}^2}{(x_S+a)^2}\approx \frac{0.247866}{S_{\lvert \lvert}}\left(\frac{r_S}{r_{nz}}\right)^2 I_{TG}.
\end{equation}
Here, $r_S$ is the radius of the skimmer. Note how all three models have the same geometrical dependencies stemming from a spherically symmetrical expansion: the $I_{TG}$ term. The intensity is then corrected upwards or downwards depending on further assumptions.

\subsubsection{Skimmer effect}\label{sec:skimmer_interference}
The intensity models discussed in the last section disregard any effect produced by the skimmer besides acting as an aperture. However, the reality is that skimmer interference is often a significant contributor to the beam's centre-line intensity \cite{palau2018center}. In its journey, a helium atom can see its trajectory perturbed by atoms backscattered from the skimmer, or more generally a perturbation of the flow caused by it.

Modelling the effect of the skimmer is a well known challenge in helium beam experiments \cite{hedgeland2005anomalous,braun1997micrometer,verheijen1984quantitative}. One of the most successful approximations to the problem is the one provided by Bird in the 1970s \cite{bird1976transition}. In this paper, Bird proposed the modified Knudsen number, and showed it to be a better predictor for skimmer interference than the Knudsen number. When designing a microscope, one should always aim for a modified Knudsen number larger than 1, as skimmer effects can decrease the intensity by as much as a factor 10 \cite{palau2018center}. The modified Knudsen number for a Lennard-Jones potential reads:

\begin{equation}
    Kn^* = Kn\left(\frac{2}{5}S_{\lvert\lvert}^2\right)^{-1/6} = \frac{1}{r_S\sigma\sqrt{2}n}\left(\frac{2}{5}S_{\lvert\lvert}^2\right)^{-1/6}.
\end{equation}
In here, the speed ratio term does not have any other effect than reducing the effective Knudsen number with respect to the normal Knudsen number. For a skimmer placed at a given distance $x_S$ from the expansion, the true dominant factor is the skimmer radius $r_S$ - smaller skimmers give larger Knudsen numbers. $\sigma$ is the scattering cross-section of the atoms and $n$ is the number density. In general, if one wants to control the optical properties of the beam, one must place the skimmer as close to the quitting surface as possible whilst having a radius that leads to a large enough modified Knudsen number. 

Since the introduction of the modified Knudsen number and the DSMC calculations by Bird there have been several attempts at modelling skimmer interference without flow dynamics simulations. One of the attempts that managed to replicate experimental data the best was a numerical model by Hedgeland et al. \cite{hedgeland2005anomalous}. In their paper, the authors propose that skimmer attenuation is mostly caused by the collision of backscattered particles with the central axis of the beam and provide a model to explain what they call an ``anomalous attenuation" of the beam at low temperatures. The advantage of this model is that the backscattered atom's cross section can be obtained from the solid angle of the beam and a series of assumptions on the nature of the He-He collisions. This improves on previous attempts that depended on parameter fitting \cite{verheijen1984quantitative}.

Although this model is very promising and replicates well the experimental data reported by \cite{hedgeland2005anomalous}, it also predicts an inverse dependency of the atom's cross section with the solid angle of the beam. This means that the cross section would decrease for bigger skimmers (that are known to produce broader beams). However measurements published in 2018~\cite{palau2018center} show that larger, equally streamlined, skimmers actually produce more interference than small skimmers - as predicted by the Knudsen number \cite{palau2018center}. These new measurements cannot be explained by Hedgeland et al.'s model. Thus, it seems that researchers are still left with no other option than to model the full interference of the beam with the skimmer if they want to obtain precise predictions of the skimmer effect. 

\subsubsection{Other effects}\label{sec:other_effects_intensity}
In addition to skimmer interference, helium atoms can interact with atoms scattered from any other element of the expansion chamber (also known as the background gas). Such interactions depend on the vacuum quality (pump capacity and in the case of a pulsed beam, size of the vacuum chamber) and can be modelled either through DSMC or through free molecular scattering. The latter is often preferred as it corresponds to a simple exponential law \cite{hedgeland2005anomalous,verheijen1984quantitative,palau2018center}:
\begin{equation}
    \frac{I}{I_S}=\exp\left(-\sigma^2n_{B_E}x_S-\sigma^2n_{B_C}a\right)
\end{equation}
Where $\sigma$ is the scattering cross-section of the atoms and $n_{B_E}$ and $n_{B_C}$ are the background number densities in the expansion chamber and subsequent chamber. These background densities should be measured by a pressure gauge far away from the beam centre line.

\section{Helium optics / Resolution limits}\label{sec:resolution_limits}
Once the centre line of the supersonic expansion has been selected by a skimmer, the helium atoms continue to travel in straight lines through vacuum until they interact with the microscope optical elements.

In this regime, the behaviour of helium atoms can be modelled by atom optics through the wave-particle duality. The wavelength, and thereby the resolution, is given by de-Broglie wavelength equation \cite{joy_choosing_2009}:
$
    \lambda_\mathrm{B}=\frac{h}{p}=\frac{h}{\sqrt{2Em}}.
$
The mass of Helium is about four orders of magnitude higher than the electron mass. Thus, at the same energy, the de-Broglie wavelength of neutral helium atoms will be two orders of magnitude smaller than of electrons and thus the potential resolutions two orders of magnitude better. According to eq. (\ref{eq:isentropic_velocity}) a room temperature helium beam has a wavelength of around 0.05~nm and an energy of around 60~meV \cite{bracco_surface_2013}. A beam cooled with liquid nitrogen and working at 120~K has a wavelength of around 0.1 nm and an energy of around 20~meV. 
However, the practical resolution limit of a Helium microscope configuration is not given by the theoretical wavelength limit, but by  aberrations and diffraction (Airy disk) broadening by the optical elements and by the signal to noise ratio in the detected signal.  As discussed in the introduction, two types of optical  elements  have been used so far to successfully produce SHeM images: Fresnel zone plates and pinholes. Fresnel  zone plates are a type of diffraction lens that focuses an incoming atomic or light beam into a small focal spot \cite{carnal1991imaging}.  Pinholes are circular openings that restrict the size of the atom beam \cite{witham_simple_2011}.

When referring to resolution, it is important to distinguish between the lateral resolution, determined by the size of the helium beam, and the ``angular resolution", given by the solid angle covered by the detector opening. The lateral resolution is what impacts the minimum feature size that can be observed and therefore is referred to in the field as ``resolution". Diffraction with detecting apertures does not degrade the lateral resolution as in light optics, because helium microscopes image by measuring the flux through the aperture and not by projecting the image onto a sensor plane. Angular resolution determines the intensity of scattered helium in a particular direction. This is mainly of relevance for contrast, in particular for 3D imaging, as multiple scattering makes it difficult to image high aspect ratio structures~\cite{lambrick_multiple_2020,lambrick2021true}.

In 2018 the concepts supra- and sub-resolution were introduced to helium microscopy \cite{fahy2018practical}. Supra resolution is the same as the lateral resolution, determined by the size of the helium beam. Sub-resolution refers to the fact that the wavelength of the helium atoms may introduce a contrast effect. This will be discussed in more detail in section \ref{sec:contrast_properties}.


The difference between the resolutions of Fresnel zone plates and pinhole microscopes is given by the contributions of  diffraction (Airy disk) and aberration terms \cite{michette1884optical}. The square of the Full Width at Half Maximum\footnote{The full width at half maximum of the beam's intensity profile \cite{palau_theoretical_2016}.} for the zone plate ($\Phi_\mathrm{ZP}$) and the pinhole 
($\Phi_\mathrm{PH}$) microscope can be written as:

\begin{eqnarray}
    \Phi_\mathrm{PH}^2 = O_\mathrm{S}^2+\sigma_A^2\qquad\quad \\
    \Phi_\mathrm{ZP}^2 = O_\mathrm{S}^2+\sigma_A^2+\sigma_\mathrm{cm}^2.
\end{eqnarray}
Where $O_\mathrm{S}^2$  indicates the geometric optics contribution to the full width half maximum. That is to say   $O_\mathrm{S}^2$ is the image of the source at the sample. For the zone plate microscope it is the demagnified image of the skimmer or the source limiting aperture~\cite{flatabo2018fast}, $\sigma_A$ is the Airy disk contribution from edge diffraction from the zone plate or pinhole and $\sigma_\mathrm{cm}$ is a chromatic aberration term that appears for the case of the zone plate.
The first equation holds under the assumption that the Fresnel number is smaller than 1, which is the case for the limit of small pinholes. For a Fresnel number larger than 1 only the geometric optics term plays a role \cite{palau_theoretical_2016}.

Besides the de-Broglie wavelength, there is no theoretical limit as to how small $O_S$ can get. However, for the case of a pinhole, $\sigma_A$ grows as $1/r_{ph}\propto 1/O_S$ \cite{palau_theoretical_2016}, where $r_{ph}$ is the radius of the pinhole, thus the Airy term increases when one tries to decrease $O_S$. For the zone plate the situation is more complex, because of the additional chromatic aberration caused by the velocity spread of the helium beam. The chromatic aberrations in a zone plate are proportional to its radius $r_{zp}$, while its Airy term depends linearly with the width of the outermost zone $\Delta r$: $\sigma_A\propto \Delta r\propto O_S$.  \cite{Salvador_theoretical_2017}. In other words, for a fix zone plate radius both the resolution and the Airy contribution decrease linearly with the same factor. This allows zone plates to reach significantly higher resolution (smaller spot size)  than pinholes \cite{Salvador_theoretical_2017,BERGIN2019112833}, see also section~\ref{sec:optimal_config}.

The resolution limits for both instruments can be explicitly obtained (see \cite{palau_theoretical_2016,Salvador_theoretical_2017}). For a pinhole microscope:
\begin{equation}
    \Phi_{PH}^{min}=K\sqrt{0.42\lambda W_\mathrm{D}\sqrt{3}}.
    \label{pinhole-resolution}
\end{equation}
Where $K=2\sqrt{2\ln 2/3}$ and $W_\mathrm{D}$ is the working distance (the distance between the optical element and the sample). The 0.42 factor comes from the Airy disk standard deviation \cite{mckechnie2016general}. In a zone plate microscope the minimum possible resolution (minimum size of the focused spot) is given by the width of the smallest zone (as the optical and Airy terms both linearly depend on $\Delta r$).
\begin{equation}
    \Phi_{ZP}^{min}=K\sigma_A\approx \Delta r.
\end{equation}
This is a well-known result from light optics for the first order focus \cite{michette1884optical}. For higher orders the focused spot size can be smaller than the width of the smallest zone~\cite{michette1884optical}. This sounds promising at first, but given that only a fraction of the beam enters into the focus (max 12.5$\%$ for the first order focus and much less for the higher orders) using a higher order focus is not an option with present detector efficiency (see section~\ref{sec:detection}). In practice, this means that the resolution is limited by nanofabrication. It is difficult to make very small free-standing zones. For this reason, experiments have been done on a so-called atom sieve zone plate configuration. The atom sieve is a zone plate superimposed with a hole pattern. The fabrication limit is now determined by how small free-standing holes can be made, rather than by how small free-standing zones can be made. In fact, the resolution limit will be even smaller than the smallest free-standing hole, because the design can be made so that a hole covers two zones and the resolution limit remains the width of a zone $\Delta r$. The idea is adapted from photonics \cite{kipp2001sharper}.  The first focusing of helium atoms using an atom sieve was done in 2015 \cite{eder2015focusing} see also \cite{flatabo2017atom}. As a final remark we can mention that it has been  shown that the Beynon Gabor zone plate performs similar to a Fresnel Zone Plate~\cite{greve2013}. The Beynon Gabor zone plate was previously cited in the literature as having a higher intensity in the first order focus that the Fresnel zone plate, however this turned out to be an artifact due to lack of sampling nodes.

Finally it should be noted that the fabrication limit of the width of the outermost zone of the zone plate introduces a minimum for $W_D$ for a given zone plate radius. From~\cite{michette1884optical} we have the following relation, where $f$ is the focal length of the zoneplate and the approximation is done under the assumption that $a$ is large so that $f\approx W_D$:
\begin{equation}
\Delta r = \lambda f/2r_{zp}  \approx \lambda W_D/2r_{zp} \rightarrow W_D=2r_{zp}\Delta r/\lambda
\label{working-distance}
\end{equation}

The discussion of resolution so far has been based on the full width at half maximum of the beams intensity profile on the sample. This is a useful working number, but does not directly correspond to the smallest feature that can be resolved. Recently Bergin et al. proposed a procedure for measuring the resolution in scanning helium microscopy using test samples with sets of slits of well-defined dimensions to establish a quantitative resolution criterion in SHeM instrumentation~\cite{Bergin2021} 

\section{Contrast properties}
\label{sec:contrast_properties}

 To the best of our knowledge the first paper dedicated to the concept of contrast in SHeM was published in 2004 by MacLaren and Allison. It discusses what contrast mechanisms are to be expected on the basis of the theory of helium scattering~\cite{MacLaren2004}.

The general theory of helium scattering has been treated in a range of books and review articles, see for example~\cite{Holst2021,Benedek2018,bracco_surface_2013, Farias1998}.  Unlike electrons, X-rays and neutrons which all interact with the core electronic cloud and atomic nuclei in the sample,  thermal helium atoms scatter off the outermost electron density distribution at the sample surface. The classical turning point for helium is a few \AA ngstroms above the surface~\cite{Alderwick2018}. 
It is no surprise therefore that the helium beam is very sensitive to surface defects such as adatoms, vacancies and atomic steps.  Experimental results on metal surfaces have shown that a defect coverage (defined as the ratio between the number of
adparticles and the number of surface atoms, both per unit area) of
$<<1\%$ of a monolayer can be detected
\cite{Poelsema83PRL,Lahee86PRL,POELSEMA19851011}. The helium
specular intensity (see below) decreases as a function of defect coverage, similarly to how a beam which crosses a gas-filled
scattering cell has its centre line intensity reduced by collisions with
gas atoms. The lost intensity turns into diffuse intensity. This analogy allows
the introduction of the concept of an effective cross section for defects. The
cross section of a single adatom as seen by helium is typically 100 $\textup{\AA}^{2}$ which exceeds by far the atomic diameter. Even for hydrogen, the cross section is estimated to be of the order of 10 $\textup{\AA}^{2}$ \cite{Kraus_2016}.
 
To understand these large cross-section values, it is necessary to analyze the scattering mechanism and in particular the helium-surface interaction potential. This interaction can be separated into a short range repulsive part, due to the overlapping of the electron densities of helium and the surface electron density, and a long range attractive part, due to the van der Waals interaction. The repulsive part taken on its own, gives a cross section of the order of the atomic size,  but including the attractive interaction which modifies the atom trajectories already far   from the surface, increases the estimated cross section value to reach the experimentally measured values. As the coverage increases, the effective cross sections of different defects start to overlap \cite{Poelsema1989book,Mete2012}.

\begin{figure}
    \centering
    \includegraphics[width=0.67\linewidth]{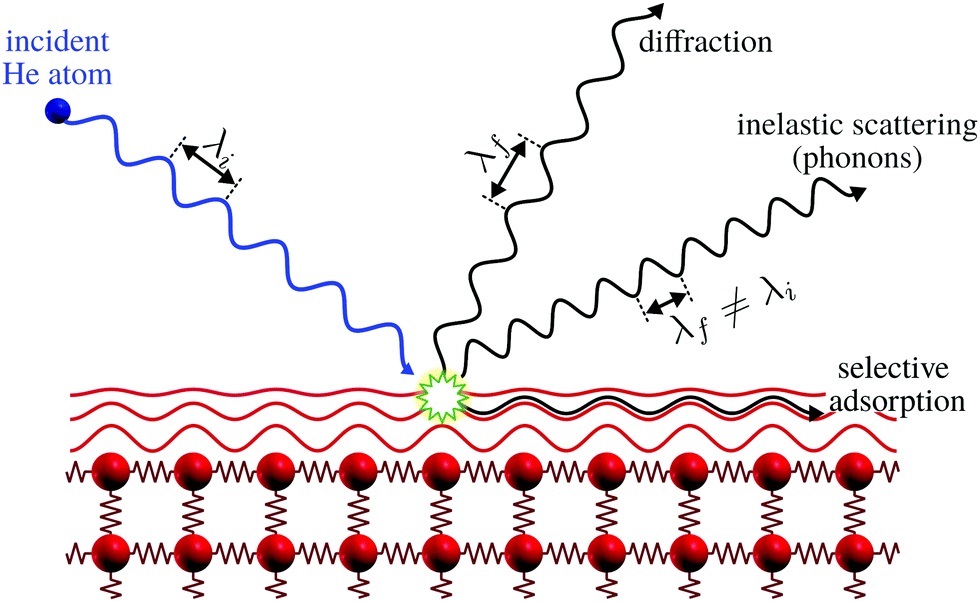}
    \caption{Illustration of the different processes for the scattering of He atoms on a surface. For crystalline surfaces diffraction is included. The helium atom scatters off the electron density distribution, indicated as red lines, without any penetration into the bulk. Selective adsorption refers to the trapping of a helium atom in the helium surface interaction potential. Here $\lambda_i$ and $\lambda_f$ denote the wavelength of the incident and scattered helium atoms, respectively. Inelastic scattering leads to a wavelength change, figure reproduced from~\cite{Holst2021}.\label{fig:heliumscattering}}
\end{figure}

The main different helium scattering processes that can occur are illustrated in Figure~\ref{fig:heliumscattering}.
The first major distinction is between elastic and inelastic scattering. In the case of elastic scattering, the energy of the helium atom is unchanged during the scattering process. In inelastic scattering an energy exchange with the surface takes place through phonon creation or annihilation. 

Specular scattering is elastic scattering, where the outgoing scattering angle is equal to the incident scattering angle. In the case where the roughness (variation in slopes) is on a length-scale bigger than the instrument resolution (focused spot size), the direction of the specularly scattered beam will vary. This is referred to in the SHeM literature as \underline{topographical contrast}. In the extreme case, when the surface is so rough on the atomic level that it acts as a perfect elastically diffuse scatterer (see section~\ref{contrast modelling}) the reflected signal will be independent from the incident beam direction, but still depend on the local, average surface normal. This is also referred to as topographical contrast.  Roughness on the atomic level can occur through the presence of atomic defects, as discussed above. In the intermediate case, where the roughness is smaller than the instrument resolution but the surface is not a perfect diffuse scatterer, the specularly reflected beam will broaden. This broadening provides a measure for the roughness variation down to the scale of the wavelength of the helium atoms (\AA ngstrom scale). This broadening effect has recently been referred to in the SHeM literature as~\underline{sub-resolution contrast}~\cite{fahy_image_2018}. First observation in a SHeM images can be found in~\cite{witham_simple_2011} It provides a unique method for fast, large area evaluation of nano-coatings~\cite{Sabrina2021,bhardwaj2021imaging}. 

As mentioned above the helium atoms have a wavelength on the \AA ngstrom scale, which is comparable to the atomic spacing in materials, so if the substrate is crystalline with a corrugated surface electron density distribution and reciprocal lattice parameters matching the k-vector component of the helium atom parallel to the surface, elastic scattering can occur in the form of diffraction. Such \underline{diffraction contrast} in SHeM was observed for the first time in 2020~\cite{bergin_observation_2020}, through imaging of a Lithium Fluoride crystal sample. Elastic scattering can also occur as resonant state scattering, also referred to as selective adsorption resonance, which occurs when the helium atom is trapped in the helium-surface interaction potential, however, this is generally a rare phenomena and has not been considered as a contrast forming process in SHeM up till now. 

Finally, and not shown in the figure, we have the case where a surface is very rough relative to the wavelength of the atoms or has a deliberately imposed high aspect ratio structure. Here the atoms may undergo more than one (elastic or inelastic) collision with the surface, which gives shadowing effects. \underline{Multiple scattering contrast} was first discussed in~\cite{witham2012increased}, where it is highlighted that the detector can be seen as being "the source of the illumination", similar to Scanning Electron Microscopy and Focussed Ion Beam imaging, where shadowing effects are also observed. Multiple scattering contrast is described in~\cite{lambrick_multiple_2020}, see also~\cite{lambrick2021true}. In the extreme case, when the atoms are thermally equibrilated with the surface through the multiple scattering, the scattering profile will be spatially similar to that of a perfectly elastically diffuse scatterer, see section~\ref{contrast modelling}.

Inelastic helium scattering has been investigated for many years using so called time of flight experiments, where the beam is chopped into short pulses and the creation and arrival time of each pulse measured, so that the time of flight (TOF) for each pulse can be converted into energy of the atoms and thus used as a measure for energy transfer with the surface - annihilation or creation of phonons. So far, however, no SHeM has been equipped with TOF.

For inelastic scattering, we distinguish between the single phonon and multi phonon regimes, also referred to as the quantum and classical regimes. In the single phonon regime, the helium atoms excite or de-excite individual phonon vibration modes. The single phonon regime occurs when single phonon annihilation or creation is the dominant inelastic process and the probability of exciting two or more phonons is small.

In the multi phonon regime several phonons are excited at the same time. This situation occurs if the vibration energies for the surface molecule charge oscillations are much lower than the energy of the incident helium atoms (the helium atoms see the surface molecules as “floppy”). In this case there will not be discrete excitations. Thermal vibrations of the surface atoms leads to an increase in multiphonon scattering with temperature. 


Inelastic scattering will lead to a loss in the elastically scattered signal. The intensity loss in the multiphonon regime $I/I_0$ is described by the Debye–Waller factor (DWF). The Debye-Waller factor was first introduced in  X-ray scattering. For helium scattering it has the form (note the temperature dependence)~\cite{barr_unlocking_2016}:
\begin{equation}\label{eq:debye}
\frac{I}{I_0}=e^{\frac{-24mT(E_icos^2\theta_{i}+D)}{Mk\Theta{_D}^2}}
\end{equation}
where $E_i$ is the incident energy of a helium atom, $m$ the mass of a helium atom, $M$ the 
surface atomic mass, $\theta_i$ the incident angle of the beam on the surface and $T$ the surface temperature  and $\Theta{_D}$ the Debye temperature, $D$ is the well depth of the helium surface interaction potential and $k$ is the Boltzmann constant. 

\begin{figure}
    \centering
    \includegraphics[width=0.67\linewidth]{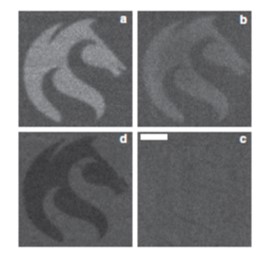}
    \caption{SHeM images showing the University of Newcastle logo in different metals on a silicon substrate. Clockwise from top left: a) gold, b) nickel c) platinum and d) chromium. Scale bar, 50 $\upmu$m from~\cite{barr_unlocking_2016}.\label{newcastle}}
\end{figure}

Equation~\ref{eq:debye} shows that inelastic scattering offers the possibility of \underline{chemical}
\underline{contrast}, since different chemical compounds on the surface will lead to different surface atomic mass, Debye temperature and well depth of the helium surface interaction potential. The first indication of chemical contrast stems from 2015 when Barr et al. published the  SHeM images shown in figure~\ref{newcastle}~\cite{barr_unlocking_2016}. They suggest that the remarkable contrast difference one observes in these images is due to the fact that different chemical elements (different metals) are being imaged. It is argued that since helium can probe subsurface resonances, chemical contrast can be provided even in the presence of multiple adsorbate layers. As an argument that the contrast is truly chemical and not sub-resolution contrast caused by differences in surface roughness, the SHeM images are compared with AFM images. It is argued that the observed SHeM contrast does not follow the root mean square roughness trend in the AFM data. One may make the remark here, that roughness is in truth a spectral density function and determined by the "ruler" used to measure it. For AFM this is the tip diameter - several nanometers, for SHeM the wavelength of the helium atoms  - less than one nm. Thus one cannot necessarily expect the roughness measured with the two methods to be comparable. 

In reality, several contrast mechanisms will often be at play  at the same time. An interesting approach for exploring contrast mechanisms is found in~\cite{bergin2019instrumentation} and~\cite{bhardwaj2021imaging}, where imaging has been done using helium beams seeded with argon and krypton. Seeded helium beams is a well established technology, used among others for thin film deposition, see for example~\cite{Torres1997}. In a microscopy context the seeded beam technique makes it possible to obtain images simultaneously with different atomic species. In principle this should make it possible to separate the different contrast mechanisms at play since, using equation~\ref{eq:debye} with different masses, $m$ and $E_i$ (atoms in a seeded beam all have roughly the same velocity\footnote{There will be a certain velocity slip depending on the mass ratio and seeding fraction.}, so these numbers can easily be evaluated). A clear contrast difference is indeed observed between the imaging with the two different gasses in~\cite{bergin2019instrumentation} and~\cite{bhardwaj2021imaging}. In practice, however, the contrast difference is not so easy to interpretate because not only the mass but also the interaction potential with the surface, and thereby the well depth, $D$ will vary. It should also be noted that the seeded beam imaging cannot be used in the zone plate configuration, since atoms with different masses at the same energy will have different de-Broglie wavelengths.

Exploration of contrast mechanisms in SHeM will no-doubt remain an intense research field in the future, just as is the case for other microscopy techniques. 

\subsection{Contrast modelling}\label{contrast modelling}

So far theoretical modelling of contrast properties has focused on topographical contrast only. Three different approaches have been used: One model assumes perfectly diffuse elastic scattering (Lambertian scattering), a second model (Knudsen flux scattering) assumes perfectly diffuse inelastic scattering, with the scattered atoms equilibrated to the surface temperature through multiple scattering. The third model assumes 
specular scattering from individual slopes. The two first models have the same spatial scattering distribution. The third model will approach the two others with increasing surface roughness.

The two first publications of theoretical methods for calculating resolutions in SHeM’s assume Lambertian reflection for modelling the scattering.~~\cite{palau_theoretical_2016,Salvador_theoretical_2017}. The term Lambertian reflection is taken from light optics, and corresponds to scattering from a perfect, diffuse reflector. The scattered lobe has a $\cos\uptheta$  spatial distribution with respect to the surface normal. The light does not change its wavelength (energy) during scattering. In other words, the scattered lobe is independent of the energy and angle of the incident beam. 

The second scattering model is Knudsen flux scattering, which has the same $\cos\uptheta$ scattering distribution as Lambertian reflection~\cite{Greenwood2002}, see also~\cite{bergin_observation_2020}. There is a fundamental difference however, between Lambertian reflection and Knudsen flux scattering. In Lambertian reflection the light does not change its wavelength (energy). This is not the case for the Knudsen flux scattering. Knudsen scattering is the scattered (desorbed) distribution, when the incident beam is totally adsorbed on the surface into the physisorption well, and then remains in the well long enough to equilibrate to the surface temperature, and then ultimately leave the surface via desorption~\cite{okeefe2001} or alternatively for rough surfaces, equilibrate to the surface temperature through multiple inelastic scattering events.

The conditions for obtaining what we refer to as Knudsen flux scattering have undergone an interesting debate. Initially, the Knudsen flux was thought to be the flux of particles that would pass through an imaginary flat plane placed in an equilibrium gas. However, this derivation was shown to be flawed by Wenaas \cite{wenaas1971equilibrium}. In 2004 Feres and Yablonsky showed that Knudsen scattering was one\footnote{But not the only one, other distributions are also possible.} of the expected results of a random billiard model for gas-surface interactions. This remains as one of the most convincing explanations for Knudsen scattering \cite{feres2004knudsen}.

The $\cos\uptheta$ distribution is used in~\cite{lambrick_ray_2018}. Here the scattering is simply labelled as diffuse scattering. It is not clear whether the Lambertian or the Knudsen flux scattering is referred to, however the result is the same as explained above.  

The last approach has been to model the scattering from rough surfaces as elastic scattering from a surface consisting of a distribution of slopes, obtained from independent AFM images~\cite{Sabrina2021}. It should be noted that in the study referred to here, the samples imaged were macroscopically flat and the explicit aim was to investigate the roughness on the (sub)-nanometer scale. A further extension is presented in~\cite{lambrick_multiple_2020} where multiple scattering is included in the modelling of images of samples with high aspect ratio.

\section{Detection}\label{sec:detection}
Detection remains the single biggest challenge in neutral helium microscopy. The big advantage of the technique - the inertness, low energy and surface sensitivity of the helium probe is its biggest disadvantage when it comes to detection. Up till now three types of neutral helium detectors have been used and/or investigated for SHeM experiments: i) Pitot-tube detectors - an accumulation (stagnation) detector, where the pressure increase from the helium flow into a small chamber is measured with a pressure gauge~\cite{eder2017zero,flatabo2018fast} (electron bombardment without mass selection). ii) electron bombardment detectors with mass selection \cite{samelin, hedgeland2007development,keppler1986neutral, alderwick2010instrumental, bullman1998development, knowling2000helium, dworski2004atom,dekieviet2000design,kalinin2006ion,bergin2019instrumentation, bergin2021low},  and iii) field ionisation detectors \cite{riley2003helium,doak2004field,doak2004-2,deponte2006brightness,piskur2008,deponte2009,o2012field}. Bolometers \cite{van1970absolute} have been used extensively in helium atom scattering experiments and photon resonance has been applied to ionize helium~\cite{hurst1975saturated,chin2012multiphoton}, but these approach have not been used in SHeM so far. 

Field ionsation detection is in principle a very attractive method, because it offers the possibility of extreme spatially resolved detection. The potential of field ionisation is demonstrated in helium ion microscopy, which uses field ionisation to generate a helium ion source, spatially confined to one ionizing atom. In an early helium microscope design proposal the sample is broadly illuminated and a mirror focuses the reflected beam onto a field ionisation detector~\cite{maclaren2002development}. So far a SHeM with field ionisation detector has not been built. One reason for this is that the field ionisation probability is strongly dependent on the velocity of the helium atoms and so would require a strongly cooled beam to achieve a reasonable detection efficiency~\cite{doak2004-2,piskur2008}

Up till now all SHeMs have used electron bombardment detectors. Helium has the highest ionisation potential of all species: around 24.6~V\footnote{ https://physics.nist.gov/PhysRefData/Handbook/Tables/heliumtable1.htm}. A key component in a helium electron bombardment detector is therefore the ioniser. Here, electrons are emitted from a negatively biased filament and accelerated by an acceleration voltage, which must be greater than 24.6~V, towards the helium beam. Positive helium ions are then created through collisions with the high-energy electrons. 

Once the helium atoms have been ionised, they need to be detected. In the simplest configuration this is done with a so-called Pitot-tube setup, used among others in~\cite{eder2017zero,flatabo2018fast} for microscope characterisation experiments. The helium beam goes through a narrow tube into a small unpumped chamber. The intensity of the helium beam is then measured by recording the pressure increase in the small volume, see~\cite{eder2013} for a description of a practical implementation. The Pitot-tube detector is very inefficient and can in practice only be used for transmission experiments, where the recorded beam intensity will be high.   

A much more efficient detection is achieved by mass separation: designed to select only those ions that interest us (helium ions coming from the beam). In SHeM (and HAS) this is often done using magnets \cite{herbert2002mass,bergin2021low} rather than the quadruple mass filters typically used in commercial residual gas analysers (mass spectrometers)~\cite{douglas2009linear}, because the magnets yield higher recorded intensities for helium. A magnet-based detector was used for the first (transmission) SHeM images~\cite{koch_imaging_2008}. A description of the design can be found here~\cite{samelin}. The helium atoms are directed from the ioniser to the mass separation stage and from the mass separation stage to the signal multiplier using ion optics \cite{bergin2019instrumentation,szilagyi2012electron,bergin2021low}. The signal multipliers used in SHeM are electron multipliers, typically tube-based multipliers known as channeltrons \cite{tuithof1975simultaneous}.


A lot of time and energy has been spent on increasing the efficiency of neutral helium detectors. Especially promising are detector systems based on solenoidal ionisers, with recent work reaching an efficiency of  as much as 0.5\%~\cite{bergin2021low} - the highest obtained to date and around three orders of magnitude higher than for the detector used in the first helium microscopy experiments~\cite{samelin}. Another promising development is a recent framework aimed at optimising the balance between signal and temporal response in neutral helium detectors. The basic idea is to use adjustable stagnation to obtain a larger helium signal, with a reported signal improvement of 27\% \cite{myles2020fast}.

\subsection{Signal to Noise Ratio}

Unfortunately a high ionisation efficiency for helium is not the only requirement for a powerful SHeM instrument. An equally crucial parameter is the signal to noise ratio, which sets a limit for the smallest signal that can be detected in a given measurement time. For electron bombardment detectors, the only detector type used in SHeMs so far, as mentioned above, there are two factors that contribute to the noise:

Firstly there is a contribution of ions from other species present in the background gas. Species such as H$_2$, H$_2$O, CO and CO$_2$ will always be present, because the vacuum is not perfect. These species all have a much lower ionisation potential than helium. Most of the ions generated will be withheld by the mass filter, but in the case of for example triple ionisation of carbon, they will be mass selected. Multiple ionisation can be strongly reduced by keeping the energy of the ionization energy as low as possible, a good vacuum also helps in general, but even so a contribution from the background gas cannot be completely prevented.

Secondly, there will be a background contribution from the helium probe itself. The part of the helium beam which is not directly reflected into the detector by the sample, will be scattered in the rest of the chamber and reflected off the walls, thus creating an additional background of helium in the chamber. A  fraction of this helium background will reach the detector. Praxis has shown that this helium background usually is the dominating contributing factor to the noise in SHeMs. The magnitude of the background will depend on factors such as pumping speed and detector opening area. It can be reduced by using a modulated beam (chopped beam) as has been demonstrated with many other techniques, however so far this has not been implemented in SHeM. This is a very hard task  since the requirement of high efficiency for the detector is generally obtained at the expenses of the response time which is lengthened, whereas modulation techniques require a relatively fast time response.This is more specifically discussed for SHeM in~\cite{witham2012increased}, see also \cite{myles2020fast}.

\section{Optimal microscope configurations}\label{sec:optimal_config}
The difficulties associated with detecting neutral helium atoms have prompted several researchers to try to optimise the design of helium microscopes to obtain a maximum beam intensity for a given resolution. 

\begin{figure}
    \centering
    \includegraphics[width=0.67\linewidth]{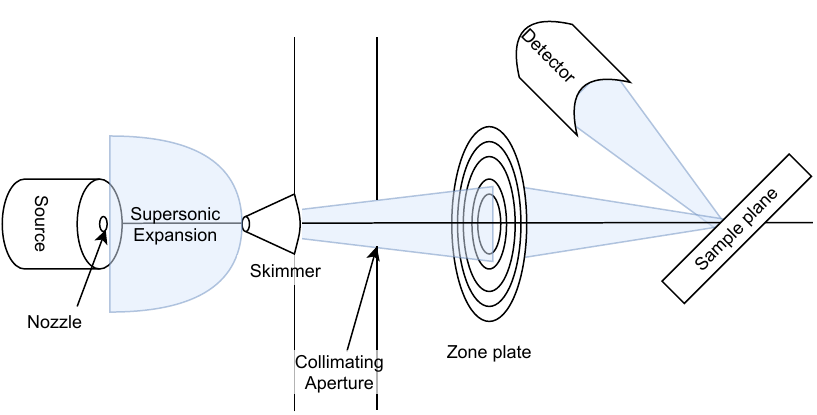}
    \caption{Simplified diagram of the zone plate microscope setup in reflection}\label{fig:zp_setup}
\end{figure}

To date, there are four papers that aim to optimise the microscope design using a theoretical framework for the beam intensity. The first paper from 2016, written by Kaltenbacher \cite{kaltenbacher2016optimization} presents an approach to optimise a microscope composed of a pinhole and two zone plates. However, Kaltenbacher does not consider the dependency of the beam centre-line intensity with the skimmer radius, rendering his approach not reliable in terms of the intensity. The next two papers from 2016 and 2018 \cite{palau_theoretical_2016,palau2018center}, by Salvador et al. present analytical approximations for calculating the optical configurations for pinhole and single zone plate SHeMs in terms of resolution and intensity. In addition, the system is also solved numerically. The zone plate configuration optimised can be found in Fig~\ref{fig:zp_setup}. The pinhole configuration has already been shown in the introduction, Fig.~\ref{fig:dastoor_pinhole_setup}.

\begin{figure}
    \centering
    \includegraphics[width=0.67\linewidth]{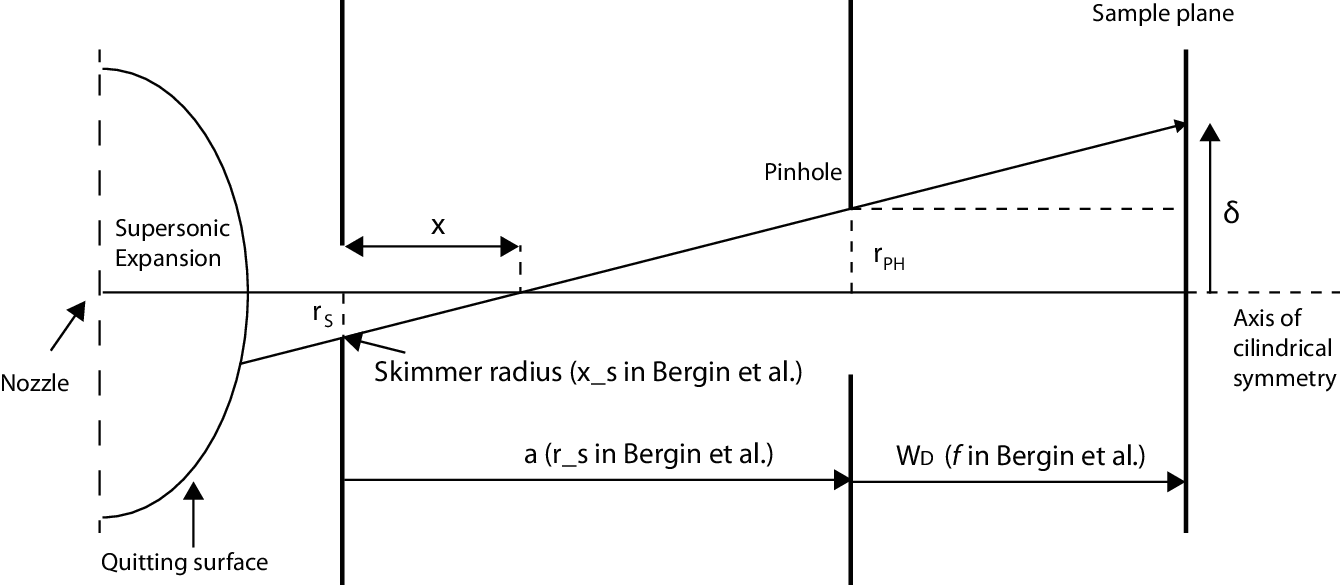}
    \caption{Sketch showing the pathway of the beam onto the sample for a pinhole microscope. The figure illustrates how the beam is limited by the skimmer in the Salvador et al. approach. The geometrical optics image of the skimmer is projected onto the
sample plane through a pinhole, giving an image of the skimmer with radius $\delta$.}\label{fig:ph_setup}
\end{figure}

The last paper on microscope optimisation by Bergin et al. from 2019~\cite{BERGIN2019112833} also presents optimisations of a pinhole and a single zone plate microscope including numerical simulations. It reaches the same qualitative conclusions on beam design as the two previous papers. The differences in the approaches of Bergin et al. and Salvador et al. are that Salvador et al. model the source using the full Sikora model (eq. (\ref{eq:I_sikora}) in this paper) whereas Bergin et al. use a simplified approximation of the Sikora model (eq. (\ref{eq:bergin_int}) in this paper). Both approaches models the beam as a Gaussian distribution with a standard derivation $\sigma$ corresponding to the skimmer radius, but Salvador et al. includes an additional limitation of the beam by the skimmer diameter as in geometrical optics, see Fig.~\ref{fig:ph_setup}.
Furthermore Salvador et al. use derivation to solve the optimisation problem, whereas Bergin et al. use Lagrange multipliers. 

Both Salvador et al. and Bergin et al. provide analytical expressions for the optimal pinhole size in a pinhole microscope configuration. Salvador et al. produce a solution valid for any working distance ($W_D$), while Bergin et al. implicitly assumes $W_D\ll a$, where  $W_D$ is the microscope working distance (distance between pinhole and sample plane, $f$ in the original Bergin et al. paper) and $a$ is the distance between the skimmer and the pinhole ($r_s$ in the original Bergin et al. paper), see Fig.~\ref{fig:ph_setup}.  

Salvador et al. obtain the following solution for the optimal pinhole radius $r^{opt}_{PH}$ (eq. 18 in the original paper~\cite{palau_theoretical_2016}).

\begin{equation}\label{eq:original_d}
 r_{PH}^{opt} = \frac{\Phi_{PH} \cdot a}{2K(a+W_D)} \stackrel{W_D\ll a}{\approx} 
    \;\frac{\Phi_{PH}}{2K}.
\end{equation}
where $\Phi_{PH}$ is the Full Width at Half Maximum of the beam at the sample plane (the resolution, see section~\ref{sec:resolution_limits}) and $K = \sqrt{8\ln(2)/3}$. We see that the diameter of the pinhole is always smaller than the Full Width Half Maximum of the beam at the sample plane. For the case $W_D\ll a$ the solution becomes independent of both $W_D$ and $a$. 

Bergin et al. obtain the following expression for the optimal pinhole diameter, $d_{PH}^{opt}$, (equation (25) in the original paper~\cite{BERGIN2019112833}): 
\begin{equation}\label{eq:bergin_vwrng}
 d_{PH}^{opt} = \sqrt{6} \Phi_{PH}^\sigma \approx 2.45 \Phi_{PH}^\sigma.
\end{equation}
Where $\Phi_{PH}^\sigma$ is the standard deviation of the helium beam at the sample plane. Since both approaches model the beam as a Gaussian distribution
we have $\Phi_{PH}=2\sqrt{2\ln2}\Phi_{PH}^\sigma$. The solution of Salvador et al. for $W_D \ll a$ in terms of Bergin et al. parameters thus becomes: 

\begin{equation}
        d_{PH}^{opt} = 
       2{\sqrt{2\ln2}\Phi_{PH}^\sigma}/{\sqrt{8\ln(2)/3}} = \sqrt{3}\Phi_{PH}^\sigma \approx 1.73 \Phi_{PH}^\sigma 
\end{equation}







For a zone plate configuration analytical solutions are harder to obtain. So far, only one has been published, obtained under several assumptions listed in \cite{Salvador_theoretical_2017}. Given these assumptions,  the optimal distance between the skimmer and the zone plate $a$, which corresponds to the solution of the following cubic equation:
\begin{multline}\label{eq:cubic}
a^3+2a^2\left(R_\mathrm{F}-\sqrt{3\Gamma}r_\mathrm{zp}\right)+aR_\mathrm{F}(R_\mathrm{F}-4r_\mathrm{zp}\sqrt{3\Gamma})\\=r_\mathrm{zp}\sqrt{3\Gamma}R_\mathrm{F}^2\left[\frac{2S^2\Phi'^2+r_\mathrm{zp}^2(\Gamma-1)}{S^2\Phi'^2-0.5r_\mathrm{zp}^2}\right].
\end{multline}

Where $\Gamma\equiv \frac{1}{3}\left(\frac{2\Delta r}{\lambda}\right)^2$ is a constant of the problem which gives the relative size of the smallest zone, $\Delta r$, of the zone plate with a given radius $r_{zp}$ compared with the average wavelength of the beam, usually $\Gamma\gg 1$. S is the speed ratio in eq. (\ref{eq:I_sikora}). $R_\mathrm{F}$ is the radius of the quitting surface and $\Phi'=\sqrt{\left(\frac{\Phi_{ZP}}{K}\right)^2-\sigma_{A}^2}$ is the corrected focal spot size (the focal spot size minus the diffraction term given by the smallest zone).

\begin{figure}
    \centering
    \includegraphics[width=0.67\linewidth]{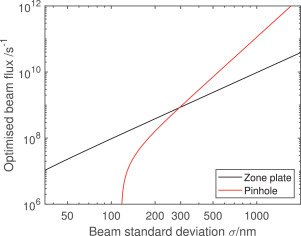}
    \caption{Plot of the optimised beam intensity (flux) versus beam standard deviation ($\sigma$) at the sample plane. $\sigma$ is a measure for the resolution, $\Phi$, which is defined in section~\ref{sec:resolution_limits} as Full Width Half Maximum of the beam at the sample plane. For a Gaussian beam we have: $\Phi=2\sqrt{2\ln2}\sigma$. We see that for lower resolutions the pinhole microscope performs better, but as the resolution improves it is outperformed by the zone plate microscope. For the configuration optimised here, based on a working distance of 1~mm and a wavelength of $\lambda=0.056~nm$, resolutions better than around 200~nm are only possible with a zone plate microscope. Figure reproduced from~\cite{BERGIN2019112833}, where details regarding the optimisation parameters can also be found.}
    \label{fig:optimisation_figure}
\end{figure}

The work done on optimal SHeM configurations has had major impact in microscope design. Most importantly, it has proven that for a given working distance (distance between optical element and sample) and given distance between skimmer and optical element, the zone plate microscope provides higher intensities at higher resolutions than the pinhole microscope (see Fig.~\ref{fig:optimisation_figure}). This is not an obvious insight, given that only around 12.5$\%$ of the beam incident on the zone plate enters the focused beam spot (see section~\ref{sec:resolution_limits}), whereas 100$\%$ of the beam that passes through the pinhole contributes to the beam spot. 

While Fig.~\ref{fig:optimisation_figure} shows that the zone plate microscope eventually "beats" the pinhole microscope for a given working distance, it is important to note that the zone plate imposes a minimum size for the working distance, see eq.~\ref{working-distance}, which is not present for the pinhole microscope. Furthermore, for the zone plate microscope to work well, an order sorting aperture needs to be inserted between the zone plate and the sample as previously discussed~\cite{eder2017zero}. In general, the smaller the distances the higher the intensity, so if the working distance is not an issue, it may be possible to conceive a pinhole design which gives higher intensity for a given resolution than a corresponding zone plate microscope. The best resolution in SHeM so far: 0.315~$\mu m$ was obtained using a pinhole microscope with a working distance of 10~$\mu m$~\cite{witham2012increased}, two orders of magnitude less than what was used in figure Fig.~\ref{fig:optimisation_figure}.

Finally it can be mentioned that the work on optimal SHeM configurations has shown that the first SHeM designs were sub-optimal. In the case of the zone plate microscope the intensity could be increased by a factor of 7 \cite{Salvador_theoretical_2017}, and in the case of the pinhole microscope by a factor 1.75 \cite{palau_theoretical_2016}. In practice the work has already led to new microscope designs using bigger skimmers and smaller skimmer-pinhole distances.

\subsection{Microscopes with micro-skimmers}\label{microskimmer}
Initial designs of helium microscopes used skimmers as small as technically feasible (a few  $\upmu\mathrm{m}$ or less). This was motivated by the desire to obtain focal spots as small as possible and micro-skimmers seemed the best way to go in the zone plate/focusing mirror set up. The first supersonic Helium beams with micro-skimmers were created by Brown et al. in 1997~\cite{braun1997micrometer}. Micro-skimmers are made by controlled drawing of glass tubes produced according to techniques developed for patch-clamp probing of cells. Brown et al. observed a broadening of the speed ratio in micro-skimmers compared to standard skimmers and suggested that this was due to geometrical imperfections and/or imperfections at the lip edge. It was recently shown that it is possible to obtain speed ratios from micro-skimmers similar to those obtained from standard skimmers~\cite{eder2018velocity}. 

Eventually it became clear that the centre-line intensity from micro-skimmers was a limiting factor for the signal intensity in the imaging spot, and a systematic study of the influence of the skimmer size on the centre-line intensity was conducted~\cite{palau2018center}, using skimmer diameters of 4, 18, 120 and 390~$\upmu\mathrm{m}$ diameters and in addition two flat apertures with diameters 5 and 100 $\upmu\mathrm{m}$. Some further measurements using a 50~$\upmu\mathrm{m}$ diameter skimmer can be found in~\cite{hustler2008aspects}. The results obtained from~\cite{palau2018center} was one of the incitements for the work on microscope optimisation discussed above. Here it is confirmed that the dependency of the centre-line intensity with the skimmer radius plays an important roles for the imaging spot intensity \cite{palau_theoretical_2016,Salvador_theoretical_2017,BERGIN2019112833}. 

Since then, new SHeM designs of both pinhole and zone plate configuration, use skimmers as big as possible given available pumping speed in combination with collimating apertures in front of the skimmer, taking into account skimmer interference at large Knudsen numbers as mentioned in Sec. \ref{sec:skimmer_interference}. This has the additional advantage that it enables fast resolution change by switching between different collimating apertures $\it{in~situ}$~\cite{flatabo2018fast}.

\subsection{Imaging with other atomic and molecular beams}

As discussed in the introduction and in section~\ref{sec:contrast_properties} imaging experiments have been carried out using other atomic and molecular beams than helium. Transmission experiments were done using deuterium~\cite{reisinger2008focus}, and reflection imaging has been done with krypton~\cite{witham_exploring_2014,bhardwaj2021imaging} and argon~\cite{bergin2019instrumentation}. In the case of~\cite{bergin2019instrumentation} and \cite{bhardwaj2021imaging} the imaging was done using so called seeded beams - a mixture of helium with the other gas. As discussed in section~\ref{sec:contrast_properties} this leads to a situation where the sample is imaged with two probes of different energies, because the atoms all move at roughly the same velocity but have different masses. As also discussed in section~\ref{sec:contrast_properties} this lead to a difference in contrast. Interestingly this strong difference in contrast was not seen in~\cite{witham_exploring_2014}, where imaging was done with two seperate beams of helium and krypton both at the same temperature. 

We here briefly discuss the advantages and disadvantages of using other substances than helium for imaging: For a given energy larger atoms (i.e. argon and krypton) have a smaller de Broglie wavelength than helium.  This is an advantage for a pinhole microscope because it reduces the diffraction at the pinhole which sets the limit for the resolution, see eq.~\ref{phdtheses}. Regarding detection, heavier atoms are easier to ionize than helium which is an advantage.  For a zone plate microscope the smaller wavelength is a disadvantage, since the zones becomes smaller and hence more difficult to fabricate. Furthermore, helium has unique properties which makes it possible to produce supersonic expansion beams with a narrower velocity distribution than what can be achieved with other atoms \cite{hillenkamp2003condensation}. A narrower velocity distribution reduces the chromatic aberrations in a zone plate microscope.  

\section{3D imaging}
Perhaps the most interesting perspective for SHeM is the potential to do true-to-size 3D imaging on the nanoscale: a nano-stereo microscope. The first 3D helium microscopy images were obtained by Myles et al.~\cite{myles2019taxonomy} in 2019. The 3D images were obtained by measuring the displacement of particular points of a 2D image when the sample was rotated by a given known angle, so called stereophotogrammetry

To avoid having to map individual sample points in different images, Lambrick and Salvador et al. in 2021  developed a theoretical framework for Heliometric Stereo, an extension of Photometric stereo to helium microscopy \cite{lambrick2021true}. The difference between stereophotogrammetry and photometric stereo is that in photometric stereo the 3D structure is recovered using variations in the intensity signal rather than geometrical displacement of the imaged points. Due to the fact that helium microscopy images are taken in an ortographic projection and constructed by imaging the sample point by point, photometric stereo can be translated to helium in an easy implementation as the image acquisition conditions are highly controlled.

Heliometic stereo is based on the fact that the intensity signal measured in detectors placed in different angles will be different and depend on the tilting angle of the imaged surface. This dependency with the scattering distribution is both a curse and a blessing: on the one hand, for heliometric stereo to be implemented straightforwardly one must know the distribution. On the other hand, however, heliometic stereo sets the perfect conditions for estimating this distribution when it is unknown as it samples it for a variety of scattering angles \cite{lambrick_multiple_2020, lambrick_ray_2018,lambrick2021true}.

\section{Conclusion and Outlook}
In this paper we present an overview of the development of neutral helium atom microscopy (SHeM) from the beginning and up to this day. New developments makes the future look promising: The exciting  perspective of true to size 3D imaging, the recent demonstration of sub-resolution contrast which allows fast characterisation of \AA ngstrom scale roughness over large areas and the improvements in detector technology, just to mention a few. All of this taken together with the developments in nanocoatings and micro and nanostructuring applications  makes it very probable that SHeM will find its use in a larger research and technology community within the next few years. The success of SHeM is likely to depend, at least to some extend, on further investigations of contrast mechanisms. Therefore, the next instrumental development step for SHeM should ideally include the possibility of chopping (pulsing) the beam so that energy resolved measurements (Time of Flight) experiments can be performed. This  would allow the different contrast contributions to be separated and analysed independently. It  would also  enable strategies for reducing the helium background, thus improving the signal to noise ratio which would increase the sensitivity of the instrument. Chopped beam experiments do however, require fast response time of the detector, which lowers the efficiency. It would also be interesting to explore contrast mechanisms further in a systematic application of beams of other atoms and helium beams seeded with these other atoms. This would allow for a systematic evaluation of the contrast mechanism according to eq.~\ref{contrast modelling}. For measurements of structures of known composition it may in some cases suffice to characterize the scattering profile independently using HAS on a flat reference surface. All new instruments should be equipped with a simple sample heating, so that water and other contaminates can be removed in order to obtain more contrast information directly from the sample material.   

\section{Acknowledgement}
We thank William Allison, Paul Dastoor, John Ellis, Daniel Farias, Holly Hedgeland, Andy Jardine, Donald McLaren and Philip Witham for useful discussions and feedback. Furthermore we thank Bruce Doak and Julian Lower with J. Peter Toennies and the Max Plank Society for  making references~\cite{Doak-22} and \cite{Lower-22-1,Lower-22-2,Lower-22-3,Lower-22-4} publicly available for the first time upon our request.  GB thanks dr. Giulia Bailo for support. BH acknowledges support of SHeM development from the European Commission through the two collaborative research projects INA, FP6-2003-NEST-A, Grant number 509014 and NEMI, FP7-NMP-2012-SME-6, Grant number 309672

\appendix

\section{Simplified model for the pinhole microscope}
During the elaboration of this comment we  found a simpler solution to the problem of optimising the set up of a pinhole microscope that does not require quadratic expressions.

We can write the following expression for an arbitrary resolution $R$:
\begin{equation}\label{eq:generalised_focal_spotsize}
    R = r_{PH} \sqrt{\psi}+r_S \sqrt{\kappa}.
\end{equation}
Comparing with our FWHM model introduced in \cite{palau_theoretical_2016} we get $\psi=(8/3)\ln (2)(1+\frac{W_D}{a})^{2}$, $\kappa=8\ln (2)(\frac{W_D}{a})^2/3$ and $R$ is the FWHM $\Phi$. 

This model generalises to other definitions of the resolution R. We note that because the square root is monotonic, we have:
\begin{equation}\label{eq:simplified_derivation}
    max(I)=max(\sqrt{I}) = max(r_{PH} r_{S}).
\end{equation}
By substitution of eq. (\ref{eq:generalised_focal_spotsize}) into eq. (\ref{eq:simplified_derivation}) we get:
\begin{equation}
    I^* = r_{PH}\left(R-r_{PH}\sqrt{\psi}\right)/\sqrt{\kappa}
\end{equation}
Where we use $I^*$ to indicate that we are dealing with a pseudo intensity that shares its maximum with $I$. Taking the derivative we get:

\begin{equation}
    R-2r_{PH}\sqrt{\psi}=0\leftrightarrow r_{PH}^{opt}=\frac{R}{2\sqrt{\psi}}.
\end{equation}

For our model, $\psi=8\ln (2)(1+\frac{W_D}{a})^2/3$ and $R$ is the FWHM. Thus, eq. (\ref{eq:original_d}) is recovered.
\section{An overview of published SHeM images and PhD thesis related to SHeM development}

Here we present an overview of, to the best of our knowledge, all SHeM images published in the scientific literature so far. The overview is presented as a chronological table (see table \ref{tab:SHeM images}). In addition we present a table of, to the best of our knowledge, all PhD theses related to the topic of SHeM (see table \ref{tab:PhD Thesis}). We have included links for download where available. Note that master theses and other student reports have not been included. We have cited PhD thesis in the main text in the cases where we have found that they contain relevant work, which has not been published in peer reviewed journals.

\begin{center}
    
	\begin{longtable}{| c | l | l |}
	\caption{Table of SHeM images published in the scientific literature so far.\label{tab:SHeM images}}\\
		\hline
		Ref. & Imaged object & Imaging beam spot size \\ 
		     & & or pinhole diameter\\
		\hline
		\hline
		\cite{koch_imaging_2008} & $\bullet$\,hexagonal copper grating (transmission) & $3\,\mu m$ and $2\,\mu m$ \\ \hline
		\cite{reisinger2008focus} & $\bullet$\,carbon holey foil (Quantifoil\textregistered, R2/1) (transmission) & $<2\,\mu m$ and $\sim2.3\,\mu m$ \\ \hline
		\cite{eder2012focusing} & $\bullet$\,carbon holey foil (Quantifoil\textregistered, R2/1) (transmission) & $<2\,\mu m$ and $\leq 1\,\mu m$ \\ \hline
		\cite{witham_simple_2011} & $\bullet$\,crushed high-field NdFeB magnet & $1.5\,\mu m$ \\
		                           & $\bullet$\,uncoated pollen grain & $1.5\,\mu m$  \\ \hline
		\cite{Witham_IEEE_simp} & $\bullet$\,aluminium sample &  \\
		                           & $\bullet$\,TEM grid, back side, with glass microspheres &   \\ \hline                       
	    \cite{witham2012increased} & $\bullet$\,uncoated \textit{Crocosmia} pollen grains & $0.35\pm 0.05\,\mu m$ \\ 
	                                & $\bullet$\,debris cluster & $0.35\pm 0.05\,\mu m$  \\ 
	                                & $\bullet$\,silicon wafer & $0.35\pm 0.05\,\mu m$  \\ \hline
	   \cite{witham_exploring_2014} & $\bullet$\,Lithium Fluoride (LiF) crystal and LiF debris & $0.35\,\mu m$ \\ 
	                                & $\bullet$\,IC test pattern, low-k dielectric on Si & $0.35\,\mu m$  \\ 
	                                & $\bullet$\,crumpled Au film on mica & $0.35\,\mu m$  \\ 
	                                & $\bullet$\,crumpled mica & $0.35\,\mu m$  \\ 
	                                & $\bullet$\,line pattern test sample, low-k dielectric on Si  & $0.35\,\mu m$  \\ 
	                                & $\bullet$\,crumpled multilayer graphene  & $0.35\,\mu m$  \\ 
	                                & $\bullet$\,\textit{Crocosmia} pollen grain  & $0.35\,\mu m$  \\ \hline
	  \cite{barr2014design} & $\bullet$\,broken copper TEM grid & $5\pm 1\,\mu m$ \\  
	                                & $\bullet$\,polymer bonded explosives & $5\pm 1 \,\mu m$  \\ 
	                                & $\bullet$\,tin spheres on carbon & $5\pm 1\,\mu m$  \\ \hline
	  \cite{fahy2015highly} & $\bullet$\,butterfly wing (\textit{Tirumala hamata})  & pinhole $\varnothing =\,5\,\mu m$ \\  
	                                & $\bullet$\,TEM grid adhered to Si wafer & pinhole $\varnothing =\,5\,\mu m$  \\ \hline
	  \cite{barr2015} & $\bullet$\,honey bee wing (\textit{Apis mellifera}) & $5.4\,\mu m$ \\  
	                                & $\bullet$\,gold logo on Si & $5.4\,\mu m$  \\ 
	                                & $\bullet$\,\vtop{\hbox{\strut gold, nickel, platinum \& chromium}\hbox{\strut logo on Si,respectively}} & $5.4\,\mu m$  \\ \hline
	  \cite{fahy_image_2018} & $\bullet$\,\vtop{\hbox{\strut hexagonal TEM grid suspended off}\hbox{\strut stainless                               steel}} & $6.9\pm 0,2\,\mu m$ \\
	                                & $\bullet$\,\vtop{\hbox{\strut central portion of a silicon nitride}\hbox{\strut x-ray window }} & $6.9\pm 0,2\,\mu m$  \\ 
	                                & $\bullet$\,\vtop{\hbox{\strut sugar crystal (\textit{sucrose})}\hbox{\strut adhered to a carbon dot}}  & $6.9\pm\,0,2\,\mu m$  \\ 
	                                & $\bullet$\,\vtop{\hbox{\strut 3D printed step sample,}\hbox{\strut resin (RSF2-GPCL-04)}} & $6.9\pm\, 0,2\,\mu m$  \\
	                                & $\bullet$\,\vtop{\hbox{\strut 3D printed angled planes,}\hbox{\strut resin (RSF2-GPCL-04)}}  & $6.9\pm 0,2\,\mu m$  \\ 
	                                & $\bullet$\,eye of a honey bee (\textit{Apis Melifera}) & $6.9\pm 0,2\,\mu m$  \\ \hline
	  \cite{lambrick_ray_2018} & $\bullet$\,TEM grid tick mark & $3.5\,\mu m$ \\  \hline
	  \cite{myles2019taxonomy}      & $\bullet$\,3D printed sample, resin (RSF2-GPCL-04) & $6.9\pm\, 0,2\,\mu m$  \\  
	                                & $\bullet$\,pyrite crystal  & $6.9\pm\, 0,2\,\mu m$  \\ 
	                                & $\bullet$\,\vtop{\hbox{\strut trichomes on Mouse-ear Cress}\hbox{\strut (\textit{A. thaliana.}) rosette leaf}}  & $6.9\pm\, 0,2\,\mu m$  \\ 
	                                & $\bullet$\,\vtop{\hbox{\strut dermal denticles on dorsal skin of}\hbox{\strut Port Jackson shark} \hbox{\strut(\textit{Heterodontus portusjacksoni})}}  & $6.9\pm\, 0,2\,\mu m$  \\ \hline
	  \cite{myles2020fast}      & $\bullet$\,silicon nitride membrane & pinhole $\varnothing =\,5\,\mu m$  \\  
	                                & $\bullet$\,\vtop{\hbox{\strut australian Emerald Tip Beetle}\hbox{\strut (\textit{Anoplognathus chloropyrus})}}   & pinhole $\varnothing =\,5\,\mu m$  \\ \hline
	  \cite{bergin_observation_2020}& $\bullet$\,cleaved LiF crystal  & pinhole $\varnothing =\, 1.2\,\mu m$ \\  \hline
	  \cite{lambrick_multiple_2020} & $\bullet$\,trenches milled into Si wafer  & pinhole $\varnothing =\,2\,\mu m$ \\ 
	                                & $\bullet$\,porous scaffold, Alvetex\textsuperscript{TM} (polystyrene)  & pinhole $\varnothing =\,2\,\mu m$ \\  \hline
	  \cite{bhardwaj2021imaging}  & $\bullet$\,MoS\textsubscript{2} films grown on SiO\textsubscript{2}/Si substrate & $23\,\mu m$\\  \hline
	  
	\end{longtable}
\end{center}  



\begin{center}
	\begin{longtable}{| l | l |}
	\caption{Table of PhD theses related to the topic of SHeM.\label{tab:PhD Thesis}}\\
		\hline
		Author name & Published, Title, link (if available) \\ 
		\hline
		\hline
		Bodil Holst & 1997, University of Cambridge, UK \\ 
		 \cite{holst1997phd} & \textit{Atom Optics and Surface Growth Studies} \\
		 & \textit{using Helium Atom Scattering} \\ 
		 & \href{https://ethos.bl.uk/OrderDetails.do?uin=uk.bl.ethos.604194}{\textcolor{blue}{ethos.bl.uk/OrderDetails.do?uin=uk.bl.ethos.604194}} \\ \hline
		Stefan Rehbein & 2001, Georg-August-University G{\"o}ttingen, Germany \\ 
		 \cite{rehbein2001entwicklung}& \textit{Entwicklung von freitragenden nanostrukturierten} \\
		 & \textit{Zonenplatten zur Fokussierung und} \\
		 & \textit{Monochromatisierung thermischer} \\
		 & \textit{Helium-Atomstrahlen} \\
		 & \href{https://cuvillier.de/de/shop/publications/3768-entwicklung-von-freitragenden-nanostrukturierten-zonenplatten-zur-fokussierung-und-monochromatisierung-thermischer-helium-atomstrahlen}{\textcolor{blue}{cuvillier.de/de/shop/publications/3768}} \\ \hline
		Donald Angus Maclaren & 2002, University of Cambridge, UK \\ 
		 \cite{maclaren2002development}& \textit{Development of a single crystal mirror} \\
		 & \textit{for scanning helium microscopy} \\
		 & \href{https://www.repository.cam.ac.uk/handle/1810/251834}{\textcolor{blue}{repository.cam.ac.uk/handle/1810/251834}} \\ \hline
		Rob T. Bacon & 2007, University of Cambridge, UK \\ 
		 \cite{bacon2007aspects}& \textit{Aspects of atom beam microscopy} \\
		 & \textit{and scattering from surfaces} \\
		 & \href{https://ethos.bl.uk/OrderDetails.do?uin=uk.bl.ethos.596236}{\textcolor{blue}{ethos.bl.uk/OrderDetails.do?uin=uk.bl.ethos.596236}} \\ \hline
		Ann Elizabeth Weeks & 2008, University of Cambridge, UK \\ 
		 \cite{weeks2008si}& \textit{Si(111) atom-optical mirrors} \\
		 & \textit{for scanning helium microscopy} \\
		 & \href{https://ethos.bl.uk/OrderDetails.do?uin=uk.bl.ethos.611911}{\textcolor{blue}{ethos.bl.uk/OrderDetails.do?uin=uk.bl.ethos.611911}} \\ \hline
		Peter Thomas Hustler-Wraight & 2008, University of Cambridge, UK \\ 
		 \cite{hustler2008aspects}& \textit{Aspects of atom-surface interactions:} \\
		 & \textit{considerations for microscopy} \\
		 & \href{https://ethos.bl.uk/OrderDetails.do?uin=uk.bl.ethos.611936}{\textcolor{blue}{ethos.bl.uk/OrderDetails.do?uin=uk.bl.ethos.611936}} \\ \hline
		Kane Michael O'Donnell & 2009, University of Newcastle, Australia \\ 
		 \cite{o2009field}& \textit{Field ionization detection for atom microscopy} \\
		 & \href{https://ogma.newcastle.edu.au/vital/access/manager/Repository/uon:6255}{\textcolor{blue}{hdl.handle.net/1959.13/802939}} \\ \hline 
		Andrew Robert Alderwick & 2010, University of Cambridge, UK \\ 
		 \cite{alderwick2010instrumental}& \textit{Instrumental and analysis tools for} \\
		 & \textit{atom scattering from surfaces} \\
		 & \href{https://ethos.bl.uk/OrderDetails.do?uin=uk.bl.ethos.608817}{\textcolor{blue}{ethos.bl.uk/OrderDetails.do?uin=uk.bl.ethos.608817}} \\ \hline
		Thomas Reisinger & 2011, University of Bergen, Norway \\ 
		 \cite{reisinger2011free}& \textit{Free-standing, axially-symmetric diffraction gratings} \\ 
		 & \textit{for neutral matter-waves:} \\
		 & \textit{experiments and fabrication} \\
		 & \href{https://bora.uib.no/bora-xmlui/handle/1956/5039}{\textcolor{blue}{bora.uib.no/bora-xmlui/handle/1956/5039}} \\ \hline
		Sabrina Daniela Eder & 2012, University of Bergen, Norway \\ 
		 \cite{eder2012neutral}& \textit{A neutral matter-wave microscope (NEMI):} \\
		 & \textit{design and setup} \\ 
		 & \href{https://bora.uib.no/bora-xmlui/handle/1956/23887}{\textcolor{blue}{bora.uib.no/bora-xmlui/handle/1956/23887}} \\ \hline
		David Matthew Chisnall & 2013, University of Cambridge, UK \\ 
		 \cite{chisnall2013high}& \textit{A high sensitivity detector} \\
		 & \textit{for helium atom scattering} \\
		 & \href{https://ethos.bl.uk/OrderDetails.do?did=1&uin=uk.bl.ethos.607777}{\textcolor{blue}{ethos.bl.uk/OrderDetails.do?uin=uk.bl.ethos.607777}} \\ \hline 
		Matthew Gordon Barr & 2015, University of Newcastle, Australia \\ 
		 \cite{barr2015imaging}& \textit{Imaging with atoms:} \\
		 & \textit{aspects of scanning helium microscopy} \\
		 & \href{https://nova.newcastle.edu.au/vital/access/manager/Repository/uon:22446}{\textcolor{blue}{http://hdl.handle.net/1959.13/1312654}} \\ \hline
		Gloria Anemone & 2017, Universidad Autónoma de Madrid, Spain \\ 
		 \cite{anemone2017development}& \textit{Development of Graphene Atomic Mirrors} \\
		 & \textit{for Neutral Helium Microscopy} \\
		 & \href{https://repositorio.uam.es/handle/10486/681667}{\textcolor{blue}{repositorio.uam.es/handle/10486/681667}} \\ \hline 
		Ranveig Flatab{\o} & 2018, University of Bergen, Norway \\ 
		 \cite{flatabo2018charged}& \textit{Charged particle lithography for the} \\
		 & \textit{fabrication of nanostructured optical elements} \\ 
		 & \href{https://bora.uib.no/bora-xmlui/handle/1956/23609}{\textcolor{blue}{bora.uib.no/bora-xmlui/handle/1956/23609}} \\ \hline
		Matthew Bergin & 2018, University of Cambridge, UK \\ 
		 \cite{bergin2019instrumentation}& \textit{Instrumentation and contrast mechanisms} \\
		 & \textit{in scanning helium microscopy} \\
		 & \href{https://www.repository.cam.ac.uk/handle/1810/290645}{\textcolor{blue}{repository.cam.ac.uk/handle/1810/290645}} \\ \hline
		Adam Joseph Fahy & 2018, University of Newcastle, Australia \\ 
		 \cite{fahy2018practical}& \textit{A practical consideration of} \\
		 & \textit{scanning helium microscopy} \\
		 & \href{https://nova.newcastle.edu.au/vital/access/manager/Repository/uon:34367}{\textcolor{blue}{http://hdl.handle.net/1959.13/1397850}} \\ \hline
		Joel Martens & 2019, University of Newcastle, Australia \\ 
		 \cite{martens2019prototype}& \textit{A prototype permanent magnet solenoidal ioniser} \\
		 & \textit{for the newcastle scanning helium microscope} \\
		 & \href{https://nova.newcastle.edu.au/vital/access/manager/Repository/uon:37871}{\textcolor{blue}{http://hdl.handle.net/1959.13/1422721}} \\ \hline
		Adrià Salvador Palau & 2021, University of Bergen, Norway \\ 
		 \cite{palau2021design}& \textit{On the design of Neutral Scanning Helium Atom} \\
		 & \textit{Microscopes (SHeM) - Optimal configurations and} \\
		 & \textit{evaluation of experimental findings} \\ \hline
	\end{longtable}
\end{center}\label{phdtheses}

 \bibliographystyle{elsarticle-num} 


\end{document}